\documentclass[useAMS,usenatbib]{mn2e}
\usepackage{amsmath}
\usepackage{amssymb}
\usepackage{graphicx}
\usepackage{pdflscape}

\title[K2VarCat II: Machine Learning Classification]{K2 Variable Catalogue II: Machine Learning Classification of Variable Stars and Eclipsing Binaries in K2 Fields 0-4}

\author[Armstrong et. al.]{\parbox{\textwidth}{D. J. Armstrong$^{1,2}$\thanks{d.j.armstrong@warwick.ac.uk}, J. Kirk$^1$, K. W. F. Lam$^1$, J. McCormac$^1$, H. P. Osborn$^1$, J. Spake$^{1,3}$, S. Walker$^1$, D. J. A. Brown$^{1}$, M. H. Kristiansen$^4$, D. Pollacco$^1$, R. West$^1$, P. J. Wheatley$^1$}
\vspace{0.4cm}\\
\parbox{\textwidth}{$^{1}$University of Warwick, Department of Physics, Gibbet Hill Road, Coventry, CV4 7AL, UK\\
$^{2}$ARC, School of Mathematics \& Physics, Queen's University Belfast, University Road, Belfast BT7 1NN, UK\\
$^{3}$Astrophysics Group, School of Physics, University of Exeter, Stocker Road, Exeter, EX4 4QL, UK\\
$^{4}$DTU Space, National Space Institute, Technical University of Denmark, Elektrovej 327, DK-2800 Lyngby, Denmark}}

\newcommand{\mytilde}{\raise.17ex\hbox{$\scriptstyle\mathtt{\sim}$}}

\begin{document}
\date{Accepted . Received}

\pagerange{\pageref{firstpage}--\pageref{lastpage}} \pubyear{2002}

\maketitle

\begin{abstract}
We are entering an era of unprecedented quantities of data from current and planned survey telescopes. To maximise the potential of such surveys, automated data analysis techniques are required. Here we implement a new methodology for variable star classification, through the combination of Kohonen Self Organising Maps (SOM, an unsupervised machine learning algorithm) and the more common Random Forest (RF) supervised machine learning technique. We apply this method to data from the K2 mission fields 0--4, finding 154 ab-type RR Lyraes (10 newly discovered), 377 $\delta$ Scuti pulsators, 133 $\gamma$ Doradus pulsators, 183 detached eclipsing binaries, 290 semi-detached or contact eclipsing binaries and 9399 other periodic (mostly spot-modulated) sources, once class significance cuts are taken into account. We present lightcurve features for all K2 stellar targets, including their three strongest detected frequencies, which can be used to study stellar rotation periods where the observed variability arises from spot modulation. The resulting catalogue of variable stars, classes, and associated data features are made available online. We publish our SOM code in \texttt{Python} as part of the open source \texttt{PyMVPA} package, which in combination with already available RF modules can be easily used to recreate the method.
\end{abstract}


\section{Introduction}
Data flows from new and planned astronomical survey telescopes are steadily increasing. This shows no sign of stopping, with LSST starting operations in \mytilde 2020. There is clearly a need for accurate, fast, automated classification of photometric lightcurves to maximise the scientific returns from these surveys. Even when later spectroscopic followup is required, finding which targets to prioritise is a necessary first step. 

The literature contains multiple examples of such classification, using a wide variety of techniques. These include a variety of supervised machine learning applications \citep[e.g.][]{Eyer:2005ce,Mahabal:2008if,Blomme:2010bq,Debosscher:2011kz,Brink:2013hv,Nun:2014kv}. Recently Random Forests (RF) have begun to gain popularity, due to their robustness and applicability to different sets of data, extracted lightcurve properties, and classification schemes \citep[e.g.][]{Richards:2011ji,Richards:2012ea,Masci:2014bk}. Several improvements have been proposed, in areas such as parametrising lightcurves with maximal information retention \citep{Kugler:2015jq}, and adjusting for training set deficiencies \citep{Richards:2011bn}. One method of \emph{unsupervised} machine learning is a Kohonen Self-Organising-Map \citep[SOM, ][]{Kohonen:1990fd} demonstrated by \citet{Brett:2004cr} in an astronomical context. Here we adopt a novel technique based on a combination of SOM and RF machine learning. SOMs can efficiently parametrise lightcurve shapes without resorting to specific lightcurve features, and RFs are capable of placing objects into classes.

In this work we apply these techniques to data from the K2 mission, the repurposed \emph{Kepler} satellite \citep{Borucki:2010dn}. K2 and its predecessor \emph{Kepler} have left a lasting mark in studies of variable stars, showing that most $\delta$ Scuti and $\gamma$ Dor stars show pulsations in both the p-mode and g-mode frequency regimes \citep{Grigahcene:2010kd}. Many studies have been performed on \emph{Kepler} variable stars \citep[e.g.][]{Blomme:2010bq,Balona:2011kw,Balona:2011gn,Debosscher:2011kz,Uytterhoeven:2011jv,Tkachenko:2013jr,Bradley:2015ep}, but few so far on K2. \citet{Balona:2015jh} studied B star variability in \emph{Kepler} and K2, and found that K2 data presented some new challenges from the original mission. Despite these, it has for example discovered the several RR Lyrae stars known outside our own Galaxy \citep{Molnar:2015tr}. \citet{LaCourse:2015jr} have also produced a catalogue of eclipsing binary stars in K2 field 0.

The initial version of this catalogue \citep{Armstrong:2015bn} classified several thousand K2 variable stars in K2 fields 0 and 1. This classification was based on an interpretation of lightcurve periodicity, and split objects into Periodic, Quasiperiodic, and Aperiodic variables. Here we improve on this initial work, by applying an automated technique to classify variables into more usual classes. We extend the classification to K2 fields 0--4, and will release updates as more K2 fields become available.

\section{Data}
\subsection{Source}
Data are taken from the K2 satellite \citep{Howell:2014ju}. K2 is the repurposed \emph{Kepler} mission, and provides lightcurve flux measurements at a 30 minute `long' cadence continuously for 80 days per target. Targets are organised into campaigns, with each campaign spanning an \mytilde 80 day period and covering several thousand objects. A much smaller number of targets (a few tens per campaign) are available at the `short' cadence of \mytilde 1 minutes. For the purposes of this work, we restrict ourselves to long cadence data only, to preserve uniformity in the data. At the time of writing, 5 campaigns had been released to the public (covering fields 0--4), with more due as the mission continues. Four of these campaigns cover \mytilde80 days, with the first campaign 0 covering \mytilde40 days. We take data for these campaigns from the Michulski Archive for Space Telescopes (MAST) website\footnote{https://archive.stsci.edu/k2/}, limiting ourselves to objects classified as stars in the MAST catalogue. This cut primarily removes a small number of solar system bodies and extended sources from the analysis. At this point, we have 68910 object lightcurves.

For the purposes of training the classifier, we also use data from the original \emph{Kepler} mission. In these cases a single quarter of long cadence \emph{Kepler} data is randomly selected. This covers \mytilde 90 days, and hence is similar to a single K2 campaign in duration and cadence. \emph{Kepler} does however have different noise properties than K2, particularly in regards to the \mytilde 6 hour thruster firing, which is present in K2 but not in \emph{Kepler}. \emph{Kepler} data was also downloaded from MAST, and the Presearch-Data-Conditioning (PDC) detrended lightcurves \citep{Stumpe:2012bj,Smith:2012ji} used.

\subsection{Extraction and Detrending}
\label{sectExtDet}
K2 data shows instrumental artefacts not previously seen in the original \emph{Kepler} mission. The strongest of these is a signal at \mytilde 6 hours, which is the timescale on which the satellite thrusters are fired to adjust the spacecraft pointing. This pointing adjustment is necessary due to drift associated with the new mode of operations, and is explained fully in the K2 mission papers. It has the unfortunate effect of causing systematic noise, due to aperture losses and inter-pixel sensitivity changes. A number of techniques have been put forward for removing this noise \citep{Vanderburg:2014bi,Aigrain:2015ew,Lund:2015cs}, including one in the previous version of this catalogue \citep{Armstrong:2015bn}. Each has advantages and disadvantages; our experience has been that while overall most techniques perform comparably, for individual objects the differences can be large. We use an updated version of our own extraction and detrending method here, which is fully described in \citet{Armstrong:2015bn}. The only change from that publication is the performing of a polynomial fit to the lightcurve, prior to detrending. This fit is performed by considering successive 0.3 day long regions of the lightcurve, and fitting third degree polynomials to 4 day regions centred on these. Outlier points more than $10\sigma$ from the initial fit are masked, and the fit redone without these points. The $10\sigma$ masking and refitting is repeated for 10 iterations. Masked points are not cut from the final lightcurves. The final fit is removed, detrending is performed, and the fit then added back in. This step was added to improve preservation of variability signals, a notable improvement on the first method. Lightcurves detrended using this method are publicly available at the MAST website.

It is important to note that, as described in \citet{Armstrong:2015bn}, our detrending method works best when performed separately on each half of the lightcurve (the exact split can be a few days from the precise halfway time). This is due to a change in the pointing characteristics of the spacecraft near the middle of each campaign, possibly the result of a change in orientation to the Sun. The precise times used to split the data are given in Table \ref{tabsplittimes}. Before conducting the analysis presented later in this work, we normalise each lightcurve half by performing a linear fit.

\begin{table}
\caption{Times of pointing characteristic change, used to split the K2 data before detrending}
\label{tabsplittimes}
\begin{tabular}{lr}
\hline
Campaign & Split Time \\
 & BJD - 2454833 \\
\hline
0 & N/A \\
1 & 2016.0\\
2 & 2101.41\\
3 & N/A \\
4 & 2273.0\\
\hline
\end{tabular}
\end{table}

With the release of campaign 3, the K2 mission team began to release its own detrended lightcurves (these are not available for earlier campaigns at the time of writing). Similarly to the other methods, we find that these perform well overall but are by no means the best choice for every object. We will apply the classifier to both our lightcurves (hereafter the `Warwick' set) and the K2 team lightcurves (hereafter the `PDC' set) for campaigns 3--4. The comparison is complicated by the fact that the above mentioned change in pointing characteristics does not occur in the usual way for these campaigns. Rather than change once in the middle of the campaign, in campaign 3 the change occurs twice, at roughly one third intervals. We do not adjust our detrending method for this, as introducing the option for another split adds an additional layer of complexity, and reduces the number of points available in each section (a risky option, as these points form the base surface used to decorrelate flux from pointing). Instead we perform the detrending with no split at all. For campaign 4, we split at time 2273 (BJD-2454833, as given in the K2 data files), and cut points up to the first change in pointing at 2240.5. This shortens each campaign 4 lightcurve by 11 days, but results in improved detrending. We do not perform such an adjustment for campaign 3 as even more data would need to be cut.

\section{Classification}

\subsection{Methodology}
We employ a classification scheme using two distinct components. These are Self-Organising-Maps (SOMs), otherwise known as Kohonen maps, and a Random Forest (RF) classifier. Each is described below.

\subsubsection{SOM}

SOMs have been tested in an astrophysical context before \citep{CarrascoKind:2014gb,Torniainen:2008cc,Brett:2004cr}, but are rarely to date applied in astronomy in practice. As such we outline their methodology here.

A SOM is a form of dimensionality reduction; data consisting of multiple pieces of information can be condensed into a pre-defined number of dimensions, and is grouped together according to similarity. In our case, the SOM takes phase folded lightcurve shapes and groups similar shapes into clusters, in one or two dimensions. The great strength of a SOM is in the unsupervised nature of its clustering algorithm. The user need not specify what groups or labels to look for; any set of similar input data, including for example previously unseen variability classes, will form a cluster in the resulting map. Similar clusters will lie near each other, those that are the same according to the input data will overlap. Furthermore, the input parameters for the algorithm are quite insensitive to small variations, making the clustering process robust \citep{Brett:2004cr}.

The key component of a SOM is the Kohonen layer. This can be N-dimensional, but we will consider 2D layers here for clarity. The layer consists of pixels, each of which represents a template against which the input data is compared. The size of the layer is unimportant as long as it is sufficiently large to express the variation present in the input data. Once trained on a set of data, the Kohonen layer becomes a set of templates, representing the observed data features that it was trained on. These templates can be examined to spot interesting features in the data set, such as variation within an already known class. Further data (or the original data itself) can be compared to the trained layer and the closest matching template found. In this way, an object is placed onto the map. 

The specific implementation of SOMs used here is described in Section \ref{sectSOMtraining}, with an example of their use shown. The result is a map against which any input K2 phase-folded lightcurve can be compared. The location of the lightcurve on the map gives us its similarity to certain shapes, such as the distinctive lightcurve of an eclipsing binary star.

\subsubsection{Random Forest}
The SOM allows us to classify and study the \emph{shape} of a given phase curve, and the sets of similar shapes found within a dataset. It does not place an object into a specific variability class. For that we utilise a RF classifier \citep{Breiman:fb}. These have been used in a number of previous variable star studies cited above. To use a RF classifier the lightcurve must be broken down into specific features, which represent the data (see Section \ref{sectdatafeatures} for those used here). These features are then paired with known classes in a training set of known variables, and the classifier fit to this set. For a given object, the RF classifier can then map sets of features to probabilities for class membership, giving the likelihood for an unclassified object to be in each class.

RF classifiers are ensemble methods, in that they give results based on a large sample of simple estimators, in this case decision trees. In this way they can reduce bias in estimation. The core components of an RF are these decision trees. See \citet{Richards:2011ji} for a concise discussion of the underlying trees and how they are constructed. The specific parameters and implementation used here are discussed in Section \ref{sectRFimplement}.

\subsection{Automated period finding}
\label{sectautoper}
Our classification methodology relies heavily on the phase-folded lightcurves of our targets. This requires knowledge of the target's dominant period. Such knowledge is available for some known variables, but not for the general K2 sample at the time of writing. As such we use the K2 photometry to determine frequencies for each target.

There are a number of methods popularly used for determining lightcurve frequencies. The most common is the Lomb-Scargle (LS) periodogram \citep{Lomb:1976bo,Scargle:1982eu}, which performs a fit of sinusoids at a series of test frequencies. Other available methods include the autocorrelation function \citep[ACF, see e.g.][]{McQuillan:2014gp} and wavelet analyses \citep{Torrence:1998wk}. We use LS here, due to its provenance and simplicity of implementation. The same arguments can be made for the ACF, which for stellar rotation periods has been shown to be more resilient than LS at detecting dominant frequencies \citep{McQuillan:2013df}. However we find removing unwanted power from frequencies and harmonics, and detecting multiple frequencies from the same lightcurve, to be simpler for the LS method, at least in the implementations that we had available. In future utilising the ACF alone or in combination with the LS may be possible.

We use the fast LS method of \citet{Press:1989hb}, with an oversampling factor of 20 run up to our Nyquist frequency of 24.5\ d$^{-1}$. To avoid excessive human interference (and maintain the `automated' status of this classification), the dominant frequencies for a target must be found without supervision. To avoid frequencies commonly associated with thruster firing noise in K2 (see Section \ref{sectExtDet}) we remove frequencies within 5\% of $4.0850$d$^{-1}$ and their $1/2$, 1st, 2nd, 3rd and 4th harmonics from the periodogram, by removing the best fitting sinusoid of form

\begin{equation}
\label{eqnLSmodel}
 z = a\sin(2\pi ft)+b\cos(2\pi ft)+c
\end{equation}

\noindent at each of these frequencies. In this model $f$ represents the frequency being removed, $t$ and $z$ the time and flux data, and $a$, $b$ and $c$ free parameters of the model. We then cut these frequencies altogether before extracting the dominant period. We also remove frequencies associated with the data cadence which commonly show power in our periodogram ($48.94355819$d$^{-1}$ and $20.394709$d)$^{-1}$) and their $1/2$ frequency harmonics, by similarly fitting and removing a sinusoid at these frequencies and then cutting the frequencies from the periodogram. We did not find it necessary to remove other harmonics of the data cadence frequencies, as doing so provided little improvement. Finally periods above 20\ d (10\ d in campaign 0) are cut, as the data baseline is not long enough to reliably determine them without the introduction of spurious noise related frequencies. At this point the most significant peak in the LS periodogram is taken.

To extract other significant frequencies, we remove the dominant frequency using a fit of the model of Equation \ref{eqnLSmodel}, then recalculate the LS periodogram, again ignoring thruster firing and cadence related frequencies as above. The remaining most significant peak is taken. To compare the power of different peaks, we calculate their amplitude $A$ using $A=\left(a^2+b^2\right)^{\frac{1}{2}}$. This is used to produce the frequency amplitude ratios used later in this work. We repeat this process to extract a total of 3 frequencies from each lightcurve.

A common weakness in period-finding algorithms occurs for eclipsing binary stars, a significant variability class. The LS periodogram often gives its highest power for half the true binary orbital period (i.e. when the primary and secondary eclipses occur at the same phase). This error is simple to spot by inspection, but harder to correct automatically. We account for this potential error source by introducing a check into the automated period finder. This phase folds each lightcurve at double its LS-determined dominant period. The phase folded lightcurve is then binned into 64 bins, and the bin values at the minimum bin and the lowest bin value between phases 0.45 and 0.55 from this minimum found. We perform two checks on these two bin values. If the initial period is correct, they should be the same. We first check for an absolute difference between the two, finding that 0.0025 in relative flux works well as a threshold. We further test that the difference between them is greater than 3\% of the range of the un-phasefolded lightcurve. We calculate this range by taking the difference between the median of the largest 30 and median of the lowest 30 flux points in the lightcurve, to avoid unwanted outlier effects. If the difference between the two tested bin values is greater than both of these thresholds, the object period is doubled. If the doubled period would be over the 20\ d upper period limit already applied (10\ d for campaign 0), the doubling is not allowed. Similar adjustments have been made in previous variability studies \citep[e.g.][]{Richards:2012ea}. Only the dominant extracted period may be adjusted in this way.

\begin{figure}
\resizebox{\hsize}{!}{\includegraphics{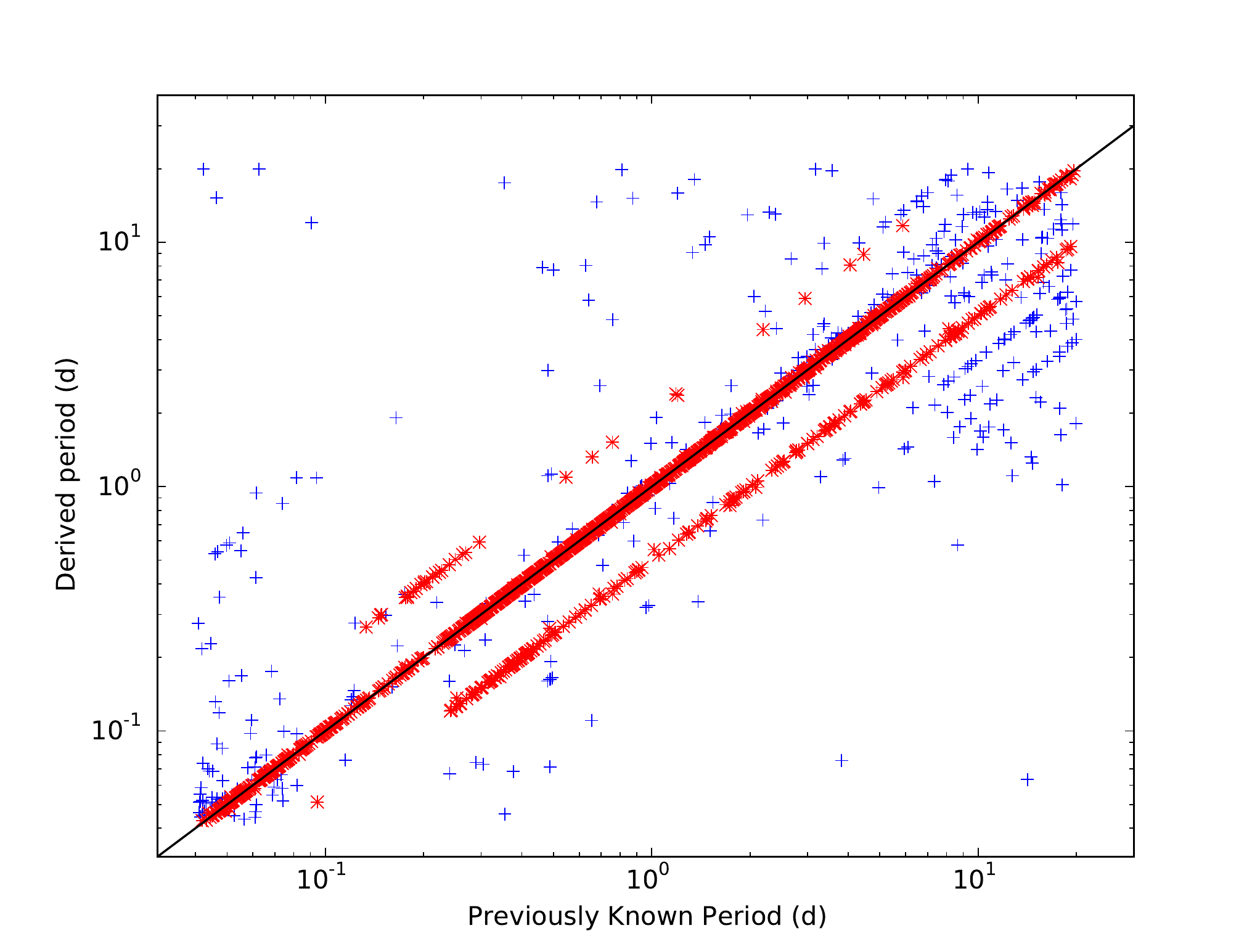}}
\caption{Periods determined using our method compared to previously known period, for a set of known variables in \emph{Kepler}. An acceptance rate of 70.3\% is obtained, 82.2\% if half and double periods are included, and 90.8\% if second and third detected frequencies are included. Variables lying at the correct, half or double frequency are plotted as red stars.}
\label{figpercomp}
\end{figure}

To test the efficacy of our automated period finding software, we trial it against a known sample of variable stars from the \emph{Kepler} data. See Section \ref{secttrainingset} for a full description of this set, which is also used a training set for our classifier. We use one randomly selected quarter of \emph{Kepler} data, to give data with a similar baseline and cadence to a single K2 campaign. There are 2128 training objects with previously determined periods (after removing objects with periods below our Nyquist period of 0.0408d). Figure \ref{figpercomp} shows the comparison between our dominant determined periods and the previously known ones. The acceptance rate is 70.3\%, rising to 82.2\% if half and double periods are included. In 90.8\% of the sample, one of our 3 determined periods finds either the previously known period or its half or double harmonic. In the remaining lightcurves, we find that either the noise obscures the known period (due possibly to different quarters with differing noise properties being used by us and previous studies) or that the dominant period has changed.

\subsubsection{Phase curve template preparation}
The SOM element of our classifier requires phased lightcurve shapes to function. We create these using the periods determined in Section \ref{sectautoper}. Each lightcurve is phase-folded on this period. For known training set objects (see Section \ref{secttrainingset}), the literature period is used. Once phase-folded, the lightcurve is binned into 64 equal width bins, and the mean of each bin used to form the phase curve that will be passed to the classifier. The exact number of bins is unimportant, as long as it gives enough resolution to see any variability in the phase curve. \citet{Brett:2004cr} used 32 bins and found satisfactory results, we use 64 as the performance decrease is small and it reduces the chances of missing rapidly changing variability such as eclipses.

It is essential that the phase curves be on the same scale and aligned, so that the classifier can spot similarities between them (see next Section). As such we normalise each phase curve to span between 0 and 1, and shift it so that the minimum bin is at phase 0. Each phase curve then consists of 64 elements, with the first being at (0,0).

\subsection{Training the SOM}
\label{sectSOMtraining}
There are variations in the literature on how precisely to train the SOM. Here we run through the procedure followed for this work. The input parameters are the initial learning rate, $\alpha_0$, which influences the rate at which pixels in the Kohonen layer are adjusted, and the initial learning radius, $\sigma_0$, which affects the size of groups. Initially each pixel is randomised so that each of its 64 elements lies between 0 and 1, as our phase curves have been scaled to this range. For each of a series of iterations, each input phase curve is compared to the Kohonen layer. The best matching pixel in the layer is found, via minimising the difference between the pixel elements and the phase curve. Each element in each pixel in the layer is then updated according to the expression

\begin{equation}
m_{xy,k,new} = \alpha e^\frac{-d_{xy}^2}{2\sigma^2}  \left(s_k - m_{xy,k,old}\right)
\end{equation}

\noindent where $m_{xy,k}$ is the value $m$ of the pixel at coordinates $x$,$y$ and element $k$ in the phase curve, $d_{xy}$ is the euclidean distance of that pixel from the best matching pixel in the layer, and $s_k$ is the kth element of the considered input phase curve. This expression is specific to 2-dimensional SOMs, but can be easily adapted for 1-dimension by setting the size of the second dimension to be 1. Note that distances are continued across the Kohonen layer boundaries, i.e. they are periodic. Once this has been performed for each phase curve, $\alpha$ and $\sigma$ are updated according to

\begin{equation}
\label{eqnsigmadecay}
\sigma = \sigma_0 e^{\left(\frac{-i*log(r)}{n_\textrm{iter}}\right)}
\end{equation}
\begin{equation}
\label{eqnalphadecay}
\alpha = \alpha_0 \left( 1 - \frac{i}{n_\textrm{iter}}   \right)
\end{equation}

\noindent where $i$ is the current iteration, and $r$ is the size of the largest dimension of the Kohonen layer. This is then repeated for $n_\textrm{iter}$ iterations. 

It is possible to use different functional forms for the evolution of $\alpha$ and $\sigma$; typically a linear or exponential decay is used. \citet{Brett:2004cr} found that the performance of the SOM was largely unimpeded by the choice of form or initial value, as long as the learning rate does not drop too quickly. We find satisfactory results for the expressions above and values of $\alpha_0=0.1$ and $\sigma_0=r$, as can be seen in the below example. The code used in this study was initially adapted from the \texttt{SOM} module of the open source \texttt{PyMVPA} package\footnote{http://www.pymvpa.org}\citep{Hanke:2009bm}, and has now been contributed as an update to that package by the authors. As such any readers wishing to use this code should look to the given reference. Note that the functional form of Equations \ref{eqnsigmadecay} and \ref{eqnalphadecay} are slightly different in the online version of the code, to preserve compatibility with older versions of the module. The formulae described here are the ones used in this work.

As an example we train a SOM on the K2 data from campaigns 0-2, as well as \emph{Kepler} data used for training the classifier (see Section \ref{secttrainingset} for a full description of the data set). We use a 40x40 Kohonen Layer. K2 data was only used if the range of variation in the phase curve before normalisation was greater than 1.5 times the overall mean of the standard deviations of points falling in each phase bin (see previous Section). This cut was imposed to avoid essentially flat lightcurves from impacting the SOM, removing \mytilde 40\% of the K2 lightcurves. The majority of these were classified as 'Noise' or 'AP' in \citet{Armstrong:2015bn}, showing that we are not removing many periodically varying sources. We note that the SOM is robust enough to work without this cut, and it is imposed only to increase the purity of the training set.

\begin{figure}
\resizebox{\hsize}{!}{\includegraphics{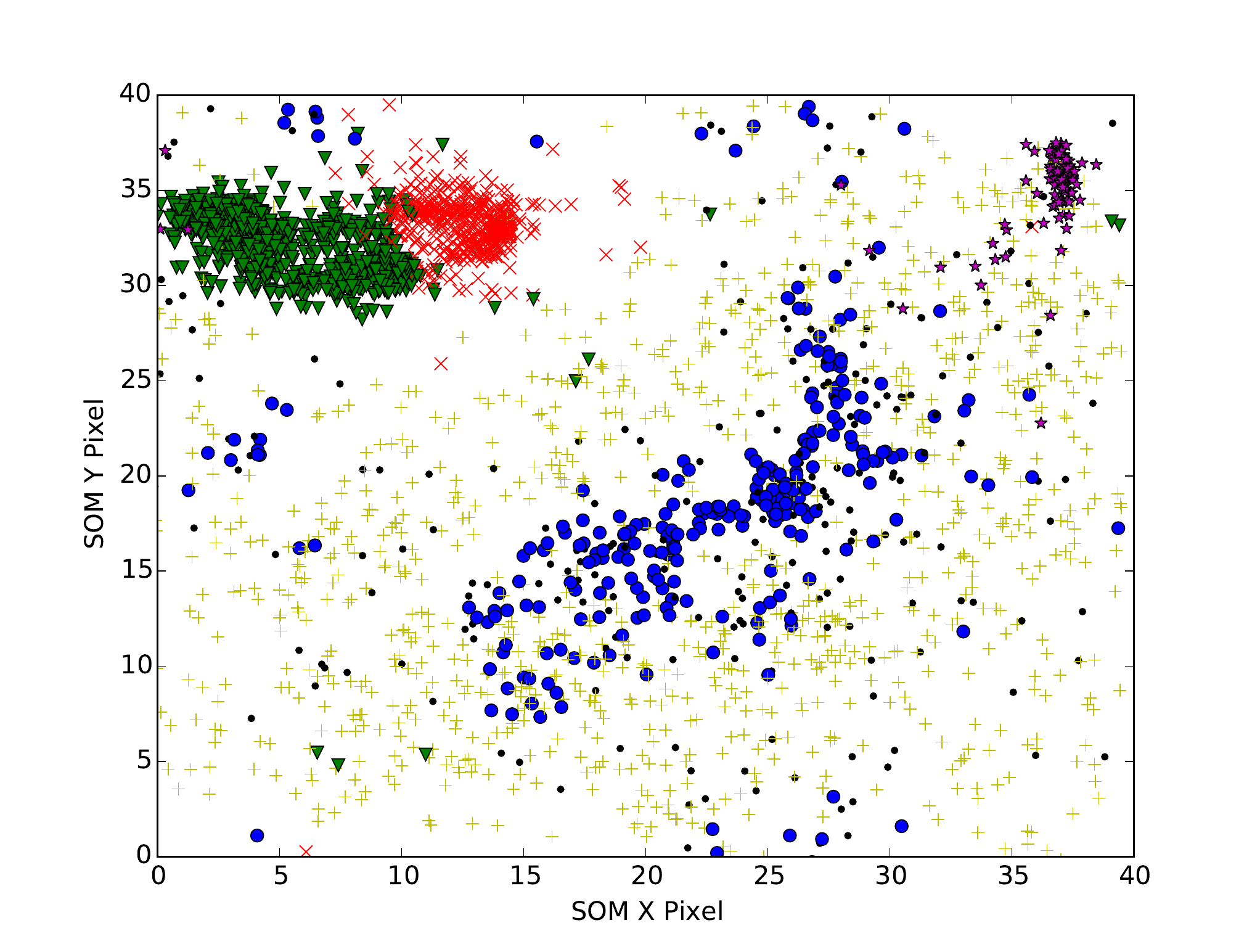}}
\caption{Known variables placed onto a SOM. Random jitter within each pixel has been added for clarity. Green triangles = 'EA' (detached eclipsing binaries), red crosses = 'EB' (semi-detached and contact eclipsing binaries), pink stars = 'RRab' (ab-type fundamental mode RR Lyraes), blue circles = 'DSCUT' ($\delta$ Scuti variables), black dots = 'GDOR' ($\gamma$ Dor variables) and yellow pluses = `OTHPER' (other periodic and quasi-periodic objects). See Section \ref{sectclassscheme} for more detail on these variability classes.}
\label{figsommap}
\end{figure}

\begin{figure}
\resizebox{\hsize}{!}{\includegraphics{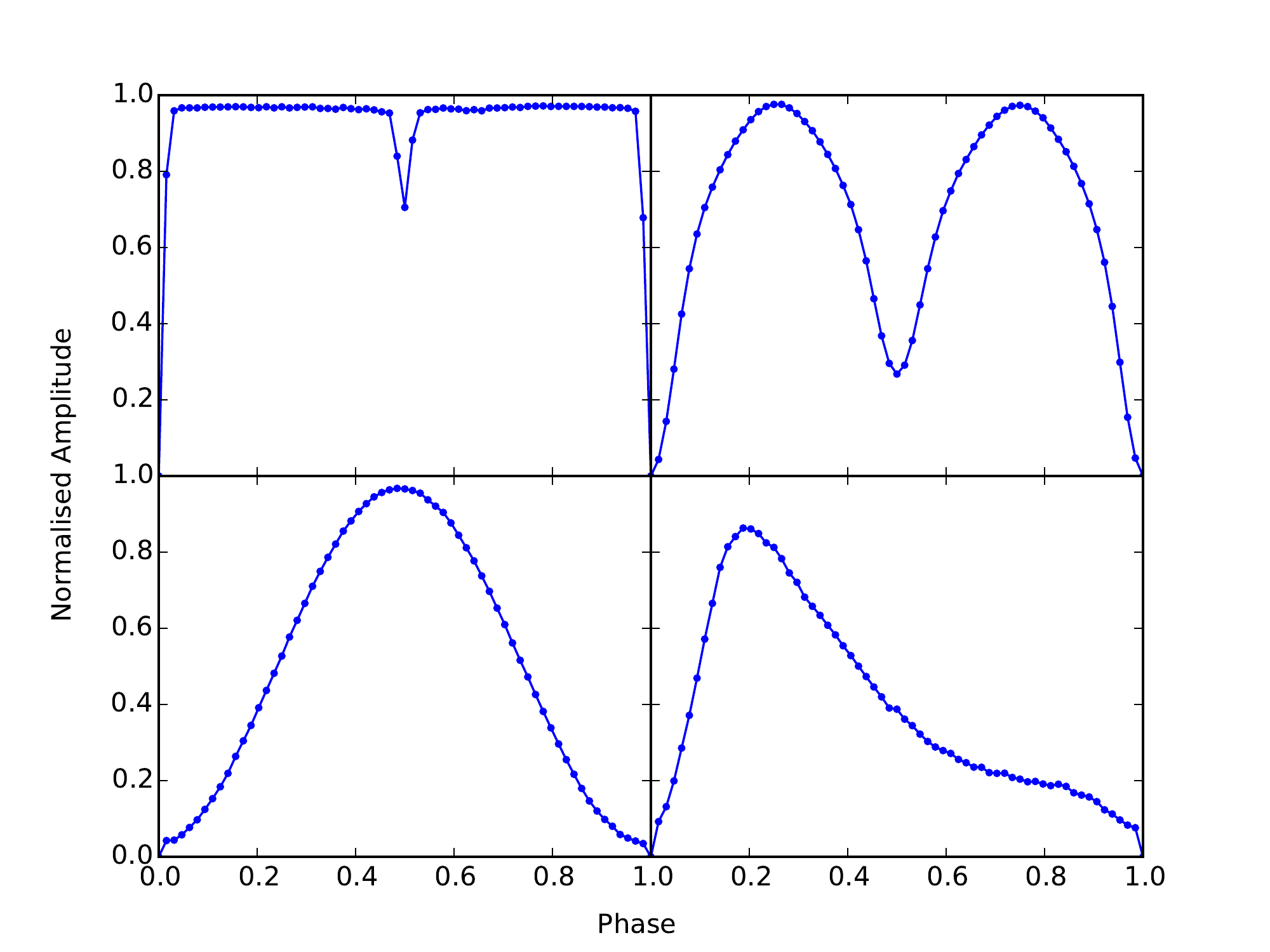}}
\caption{Template phase curves from the Kohonen layer of the SOM in Figure \ref{figsommap}. Clockwise from top left, templates are for pixel [13,34] (EA), [6,32] (EB), [37,35] (RRab) and [25,19] (DSCUT). See Section \ref{sectclassscheme} for a description of the classes. Note that templates do not have to span the range 0-1, even if the input phase curves do. Note also that all these templates were found from initially random pixels without any human guidance or input.}
\label{figsomtemplates}
\end{figure}

We take the known \emph{Kepler} variables, along with `OTHPER' other periodic and quasi-periodic objects from K2, and plot them on the resulting SOM in Figure \ref{figsommap}. Clear groups can be seen, with eclipsing binary types well differentiated but bordering each other, as would be expected. RR Lyraes are very well grouped, and $\delta$ Scuti variables cluster but more weakly. Example templates from the Kohonen layer are shown in Figure \ref{figsomtemplates}, representing the major clusters seen. Note that the size of a group is determined by a number of factors, including the number of input objects matching it, and the extent of small variations within the group. As there are many more sinusoidal variables than eclipsing binaries or RR Lyraes, the $\delta$ Scuti, $\gamma$ Doradus and `OTHPER' groups fill most of the map. Different regions within these groups show for example slight skews from a pure sinusoid, and may represent interesting intra-class differences. $\delta$ Scutis lying near the eclipsing binary groups have likely been mapped using double their true period, and so look similar to a contact binary star. They may also have been previously misclassified. It is also interesting to see that $\delta$ Scutis and `OTHPER' objects overlap, as would be expected given that their phase curve shapes are not particularly distinctive to their respective classes. `OTHPER' objects also overlap with the RR Lyrae cluster, and likely mark out newly discovered RR Lyrae stars.

The SOM used for final classification is the same as that described above, but using only one dimension of 1600 pixels. This produces the same clustering results, but is less useful for visualisation. We use only one dimension so that the other part of our classifier (the RF) can more easily make use of the information contained within the SOM.

\subsection{Data Features}
\label{sectdatafeatures}
For the classification of variables into classes, we use a number of specific features of each lightcurve. This is common practice in general classification problems \citep[e.g.][]{Richards:2011ji}. However, there is a subjective element to selecting features, and it can be desirable to minimise this if possible \citep[see e.g.][]{Kugler:2015jq}. We do so through the use of the SOM. This encodes the shape of the phase curve into one parameter (the location of the closest pixel in the SOM to the lightcurve in question), rather than a series of features, none of which may capture the desired shape properties.

There are however other features which are useful and which are uninformed by the SOM. A key example is the dominant (most significant) period of the lightcurve. Other significant frequencies can also be used, and in some cases many more have been studied. We only use the three most significant periods here.

The full range of features used is described in Table \ref{tabfeatures}. These features are incorporated largely to separate out lightcurves which show purely noise, something which is generally uninformed by the SOM, as well as those without one particularly dominant frequency. We take the potentially controversial step of adjusting some of the noise related features between the \emph{Kepler} and K2 datasets, due to the differing noise properties between each set. This is unavoidable here, as the scatter and increased noise in K2 causes catastrophic errors in the classifier if \emph{Kepler} lightcurves are used as they come. In this case the general result is that the vast majority of K2 objects are classified as Noise. This problem is solved by multiplying the marked features in Table \ref{tabfeatures} by a factor to align their median values with those of K2. These features are those driven primarily by dataset noise, rather than those associated with periodicity (noise-related periodicity is assumed to have been removed by the procedure in Section \ref{sectautoper}). As the \emph{Kepler} data used all comes from known variable stars, the median of the features is not strictly comparable to K2, where the data comes from the whole target list. As such we set the multiplication factor so that the median of the non-eclipsing binary \emph{Kepler} data features is increased to equal the median of the `OTHPER' K2 data features. Eclipsing binaries are left alone, as their features are in our case dominated by the binary eclipses.

A similar problem arises when studying the PDC lightcurves. These have different characteristics to the Warwick lightcurves. Assuming that the intrinsic distribution of stellar variability should be the same across fields, this difference is due to the differing detrending methods. We adjust for it in the same way and to the same features as above, marked in Table \ref{tabfeatures}. As we do not have prior classifications for fields 3--4, the factor is applied to the whole dataset, and set so as to match the medians of these features between the PDC campaigns 3 and 4 and the Warwick campaigns 0--2. Each PDC campaign is adjusted separately.

It would be desirable to use colour information as a feature to aid classification of variability types connected to specific stellar spectral types. However, colours are not uniformly available for the K2 sample, although some can be found through a cross-match with the TESS input catalog \citep{Stassun:2014wz}. As such we do not use them, as doing so would mean large fractions of the K2 targets would need to be disregarded. This has consequences for the variability classes we use, see Section \ref{sectclassscheme}.

\begin{table}
\caption{Data Features}
\label{tabfeatures}
\begin{tabular}{ll}
\hline
Feature Name & Description \\
\hline
period & Most significant period (Section \ref{sectautoper})\\
amplitude & Max - min of phase curve			\\
period\_2  & Second detected period (Section \ref{sectautoper}) \\
period\_3 & Third detected period (Section \ref{sectautoper}) \\
ampratio\_21 & period\_2 to period amplitude ratio \\
ampratio\_31 & period\_3 to period amplitude ratio \\
SOM\_index & Index of closest pixel in 1D SOM		\\
SOM\_distance & Euclidean distance to closest pixel 			\\
    &  in 1D SOM  \\
p2p\_98perc $^a$& 98th percentile of point to point scatter\\
                     &   in lightcurve			\\	
p2p\_mean $^a$&   Mean of point to point scatter in lightcurve                               \\
phase\_p2p\_max & Maximum point to point scatter in binned\\
                              &    phase curve			\\
phase\_p2p\_mean & Mean of point to point scatter in binned \\
                                & phase curve			\\
std\_ov\_err $^a$& Whole lightcurve standard deviation over \\
                     &mean point error			\\
\hline
\multicolumn{2}{l}{$^a$ adjusted between datasets, see text.}
\end{tabular}
\end{table}

\subsection{Classification Scheme}
\label{sectclassscheme}
An important decision is in which variability classes to use. We experimented with classifying RR Lyrae (subtype ab), $\delta$ Scuti, eclipsing binary (split into detached, subtype EA, and semi-detached or contact, subtype EB), $\gamma$ Dor, and so-called ROT variables, a class applying to likely rotationally modulated lightcurves seen in \citet{Bradley:2015ep}. We also attempted to split the $\gamma$ Dor class into symmetric, asymmetric, and 'MULT' classes, as defined in \citet{Balona:2011kw}. This approach had varied success; RR Lyrae ab, $\delta$ Scuti, $\gamma$ Dor and eclipsing binary classes performed well, but we found that the $\gamma$ Dor subtypes were not well constrained by our available features. This may be because we lack sufficient training objects to reliably map the range of features offered by these subtypes. This problem could be navigable when an increased sample of objects is available through K2, and we plan to address this in later work. 

Similarly, we found that the 'ROT' class was not very coherent - the classifier struggled to identify regions in parameter space corresponding to these variables. This likely arises due to the tendency of this class to have an indistinct cluster of low frequency peaks rather than one clear signal \citep{Bradley:2015ep}. Rather than use the ROT class by itself, we make use of the previous version of this catalogue, which contained a `QP' quasiperiodic variable class. This class contains a number of variable types, but is characterised by periodic variability that is not strictly sinusoidal, and changes in amplitude and/or period. We use this as a variable classification, to catch interesting variables of astrophysical origin which are not one of the five other classes (RR Lyrae ab, EA, EB, $\delta$ Scuti, $\gamma$ Dor). It is likely dominated by spot-modulated stars, but also contains other variables such as Cepheids. We rename this class to `OTHPER' for `other periodic' to avoid confusion, as variables which are strictly periodic but not in another class can be classified by this group.

We considered including other variable classes, such as Cepheids, the other RR Lyrae subtypes (first-overtone or multimode RR Lyraes), and Mira variables. We could not find sufficient training set objects in any of these classes (less than 20 in each case). While it is possible to attempt classification with small training sets, rather than present a weak or unreliable classification for these classes we prefer to wait for more K2 data. As more fields are observed, more training set objects will become available. We intend to include more classes in future versions of this catalogue.

Finally, we include 'Noise', non-variable lightcurves, as a class label. This leave 7 classes, DSCUT ($\delta$ Scuti), GDOR ($\gamma$ Doradus), EA (detached eclipsing binaries), EB (semi-detached and contact eclipsing binaries), OTHPER (other periodic and quasi-periodic variables), RRab (RR Lyrae ab type) and Noise. It is important to note that as we do not have colour information, there will be degeneracy in the DSCUT class between true $\delta$ Scutis and $\beta$ Ceph variables, as in \citet{Debosscher:2011kz}. This is also true for slowly pulsating B stars, which are degenerate with $\gamma$ Dor variables.

\subsection{Training Set}
\label{secttrainingset}
Although the SOM described is unsupervised and so requires no training set, the RF classifier we use for final classification does. An ideal training set would consist of a set of known variable stars from the K2 mission, to which we can fit the classifier. Some previous classification work on K2 has been done (for B stars \citep{Balona:2015jh}, for eclipsing binaries \citep{LaCourse:2015jr}, and in the previous version of this catalogue). These sources however suffer from either small numbers, only being applicable to a few variable types, or in the \citet{Armstrong:2015bn} case using variability classes derived from the lightcurves rather than externally recognised types. We cross matched the observed K2 targets in fields 0--3 (4 was not available at that time) with catalogues of known variable stars, including those from AAVSO\footnote{www.aavso.org}, GCVS \citep{Samus:2009tf} and ASAS \citep{Richards:2012ea}. This led to a small number of targets (a few tens of each class at best), not enough for a full training set. As such, we turned to the original \emph{Kepler} mission. Much classification work has been done on the \emph{Kepler} lightcurves. The data has differing noise properties to K2 data, but the same cadence, instrument, and if only one 90 day quarter of data is used a similar baseline to a K2 campaign.

Although multiple works are available offering classified variable stars in \emph{Kepler}, we limit ourselves to a small number of relatively large scale catalogues, in order to maintain homogeneity among classification methods and simplify the process. We began by taking the EA, EB, DSCUT classes from \citet{Bradley:2015ep} We also took ROT, SPOTM and SPOTV, low frequency variables likely due to rotational modulation, reclassifying these objects as OTHPER. We supplemented the DSCUT set with those from \citet{Uytterhoeven:2011jv}. The bulk of our eclipsing binary training set come from the Kepler Eclipsing Binary Catalogue \citep{Prsa:2011dx,Slawson:2011fg}. We removed all heartbeat binaries \citep{Thompson:2012ca} and those where the primary eclipse depth was less than 1\%. A threshold of 1\% was implemented in order to avoid shallow, likely blended binary eclipses from being included in the training set and hence increase training set purity. This also avoids the problem of noisy lightcurves with instrumental systematics of order a percent being misclassified as eclipsing binaries. Binaries were then classified as EA or EB based on a morphology threshold of 0.5 (see \citet{Matijevic:2012di} for a discussion of morphology in this context). For RR Lyrae stars we use the list in \citet{Nemec:2013bp}. Fundamental mode subtype ab stars were labelled RRab, and the first-overtone subtype c stars classified as OTHPER. To increase this relatively small RR Lyrae sample we used the results from the K2 AAVSO cross-match, taking fundamental mode RR Lyraes and adding them to the RRab training set. The B-star catalogue of \citet{Balona:2015jh} was also used, with the SPB class reclassified as GDOR (given the degeneracy between GDOR and SPB present without temperature information) and the ROT class being reclassified as OTHPER.

For the OTHPER and Noise classes, we also use our previous catalogue. This contained 5445 OTHPER (QP in the original catalogue) and 29228 Noise objects in fields 0--2, with labels assigned by human eyeballing. To avoid having an excessive disparity between training set classes, we downsample this set to 1000 of each class, selected randomly, which are then added on to the \emph{Kepler} OTHPER set above. This also makes the results on fields 0--2 more independent, as we can compare previously classified OTHPERs (the majority of which are now not in the training sample) with newly found ones. To reduce the impact of potential mistakes in the previous catalogue, we removed the small number of objects in the OTHPER training set which were in an initial run of this classifier reclassified as another class. Objects with a probability of being in the RRab class of greater than 0.2 were also removed, as the probabilities for the RRab class are not well calibrated (see Section \ref{sectprobcal}). These cuts caught \mytilde 50 objects misclassified as OTHPER and \mytilde 30 objects misclassified as Noise out of the 1000 each initially selected. 

The final classes and number of objects in each training set are shown in Table \ref{tabtrainingset}.

%key question - avoiding the noise

\begin{table}
\caption{Training Set}
\label{tabtrainingset}
\begin{tabular}{lr}
\hline
Class & N objects \\
\hline
RRab & 91\\
DSCUT & 	278	\\
GDOR & 233 \\
EA & 	694	\\
EB &   759	\\
OTHPER &  1992 \\
Noise &      976         \\
\hline

\end{tabular}
\end{table}

\subsection{Random Forest Implementation}
\label{sectRFimplement}
We use the implementation of RFs in the \texttt{scikit-learn} \texttt{Python} module\footnote{http://scikit-learn.org/stable/}. There are several input parameters for an RF classifier. The key ones are the number of estimators, the maximum features considered at each branch in the component decision trees, and the minimum number of samples required to split a node on the tree, which controls how far each tree is extended. In a typical case, increasing the number of estimators always leads to improvement in performance but with decreasing returns and increasing computation time. The theoretical optimum maximum features for a classification problem is the square root of the total number of features, in our case 3. We optimise the parameters using the 'out-of-bag' score of the RF. When training, the classifier uses a random subset of the total data sample given to it for each tree, to reduce the chance of bias. The left out data is then used to test the performance of the tree -- its known class is compared to the predicted class, giving a performance metric between 0 (for absolute failure) and 1 (for perfect classification). Maximising this metric allows us to optimise the parameters. We find the best results for 300 estimators, a maximum of 3 features, and 5 samples to split a node. These parameters are used for classification. Additionally we apply weights to the training set, so that each class is inversely weighted according to its frequency in the training set (input option class\_weight=`auto'). This makes sure that classes with more members (such as OTHPER and Noise) do not drown out other classes, and in effect imposes a uniform prior on the class probabilities.

There are several random elements in our method. These are the selection of the OTHPER and Noise training sets, as well as certain elements of the RF. Random subsets of training objects and features are selected for each decision tree as part of the RF method, to avoid bias. To minimise any effects of this randomness (especially the OTHPER and Noise selection), we train 50 classifiers with the above parameters and repeat the selection for each, applying each classifier to the K2 dataset. The average class probability across the classifiers gives the final result.
 
\begin{figure}
\resizebox{\hsize}{!}{\includegraphics{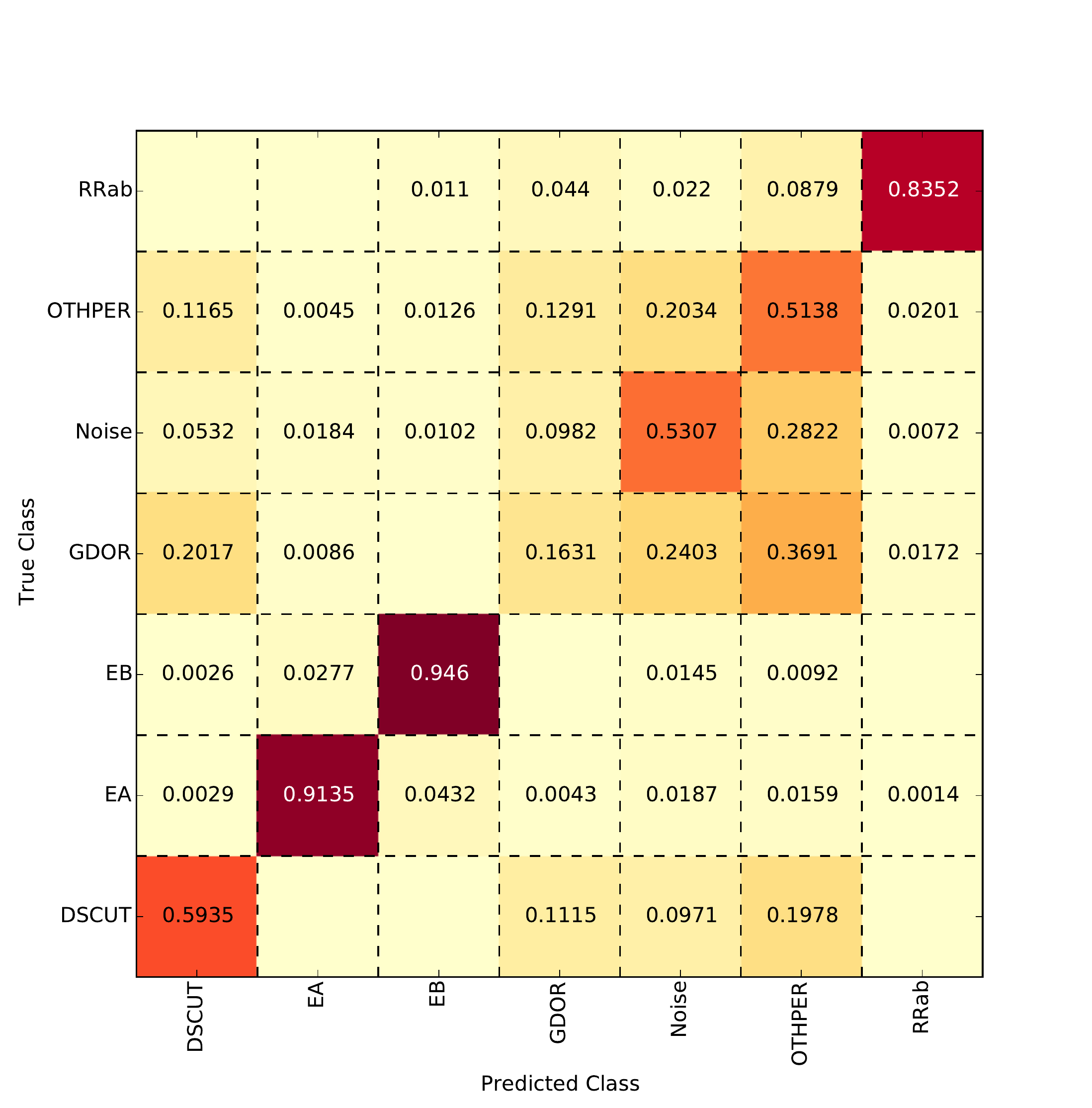}}
\caption{Confusion matrix for a RF considering only SOM map location, generated using leave-one-out cross validation. Text shows the percentage of each sample which was classified into the relevant box. Correct classification lies on the diagonal.}
\label{figconfmatrixsom}
\end{figure}

To explore the power of the SOM method, we trial the RF on only the SOM map location (SOM\_index). The classifier is cross-validated by taking one training set member and training the classifier on the remaining members (so-called leave-one-out cross validation). The left out object is then tested on the classifier, and the process repeated for each member. The performance of the classifier is best described by a `confusion matrix', shown in Figure \ref{figconfmatrixsom}. This shows what proportion of training members in each class were assigned to which other classes. In the ideal case each object is predicted correctly. Here we can see clearly which classes are well-informed by the SOM. RRab, EA, and EB classes are strongly recovered, as expected from their strong localisation in Figure \ref{figsommap}. The DSCUT class is also recovered although less so. On the other side, OTHPER and Noise classes are found more weakly, and GDOR barely at all, due to the often multiple pulsation frequencies in this class combining to produce no distinctive phase curve shape. This demonstrates the power of the SOM alone to classify certain classes of variable stars. 

\begin{figure}
\resizebox{\hsize}{!}{\includegraphics{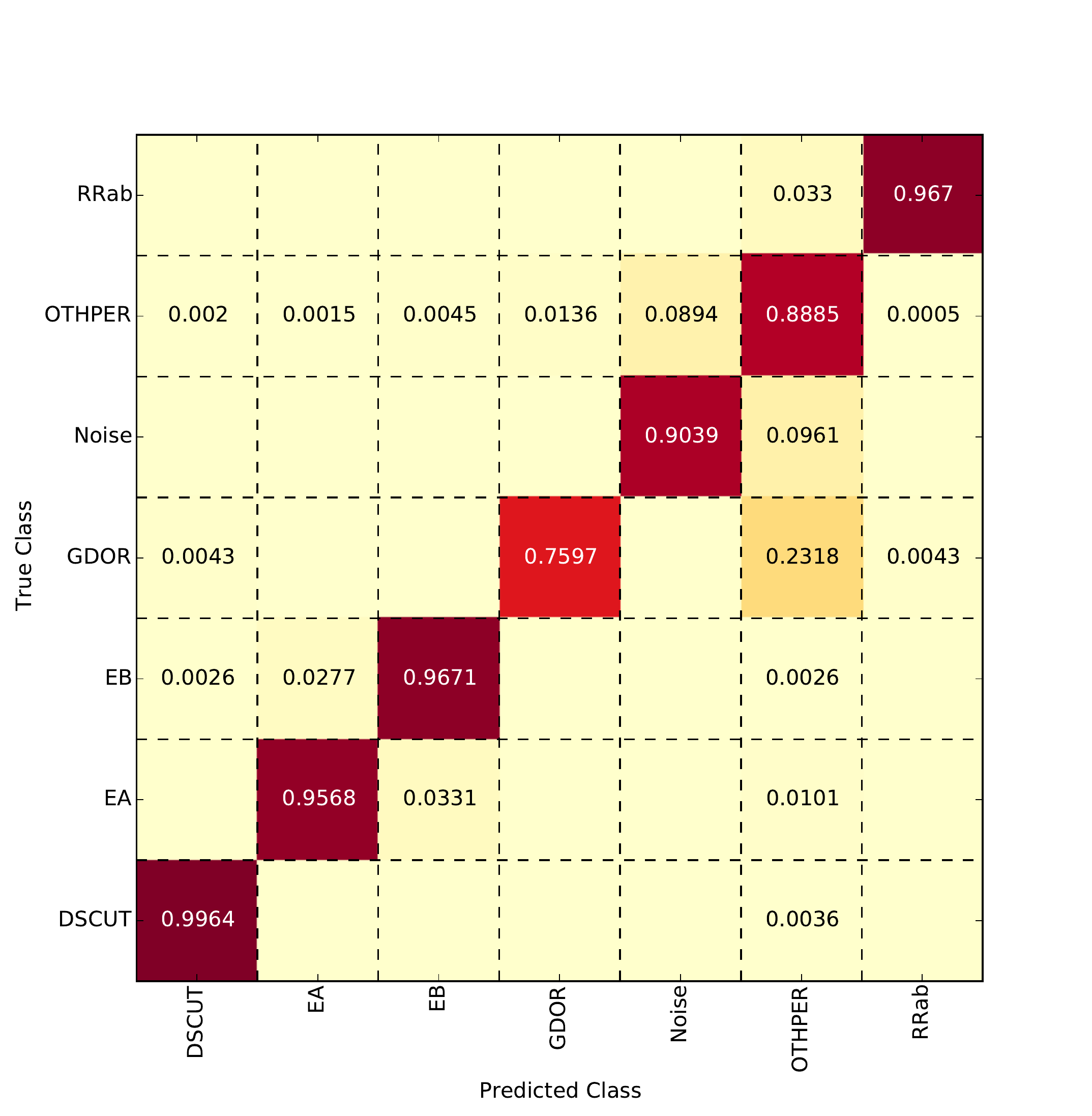}}
\caption{Confusion matrix for a RF considering all features and classes, generated using leave-one-out cross validation. Text shows the percentage of each sample which was classified into the relevant box. Correct classification lies on the diagonal.}
\label{figconfmatrix}
\end{figure}

Moving on to the full classification scheme, we test the RF in a similar manner. All 7 classes are used, and the classifier cross-validated as before. The resulting confusion matrix is shown in Figure \ref{figconfmatrix}. It highlights some interesting cases. Firstly, the classifier works well, with an overall success rate of 92.0\%. There is some porosity between the two eclipsing binary classes, with objects of one class being placed into the other. As there is no rigid boundary in lightcurve shape between them, this is to be expected. Similarly there is some spread between OTHPER and Noise. This is not desirable, but the numbers involved are low, and represent objects with either variability only just emerging above the noise or objects with unusual noise properties. The biggest misclassification occurs between the GDOR and OTHPER classes. This arises due to the less distinct nature of the OTHPER class - it acts as a `catch-all' class to find any periodic or quasi-periodic variables which do not fit the other classes. GDOR objects can in some circumstances present similar lightcurve features to for example fast rotating stars, leading to some confusion between the classes.

\begin{figure}
\resizebox{\hsize}{!}{\includegraphics{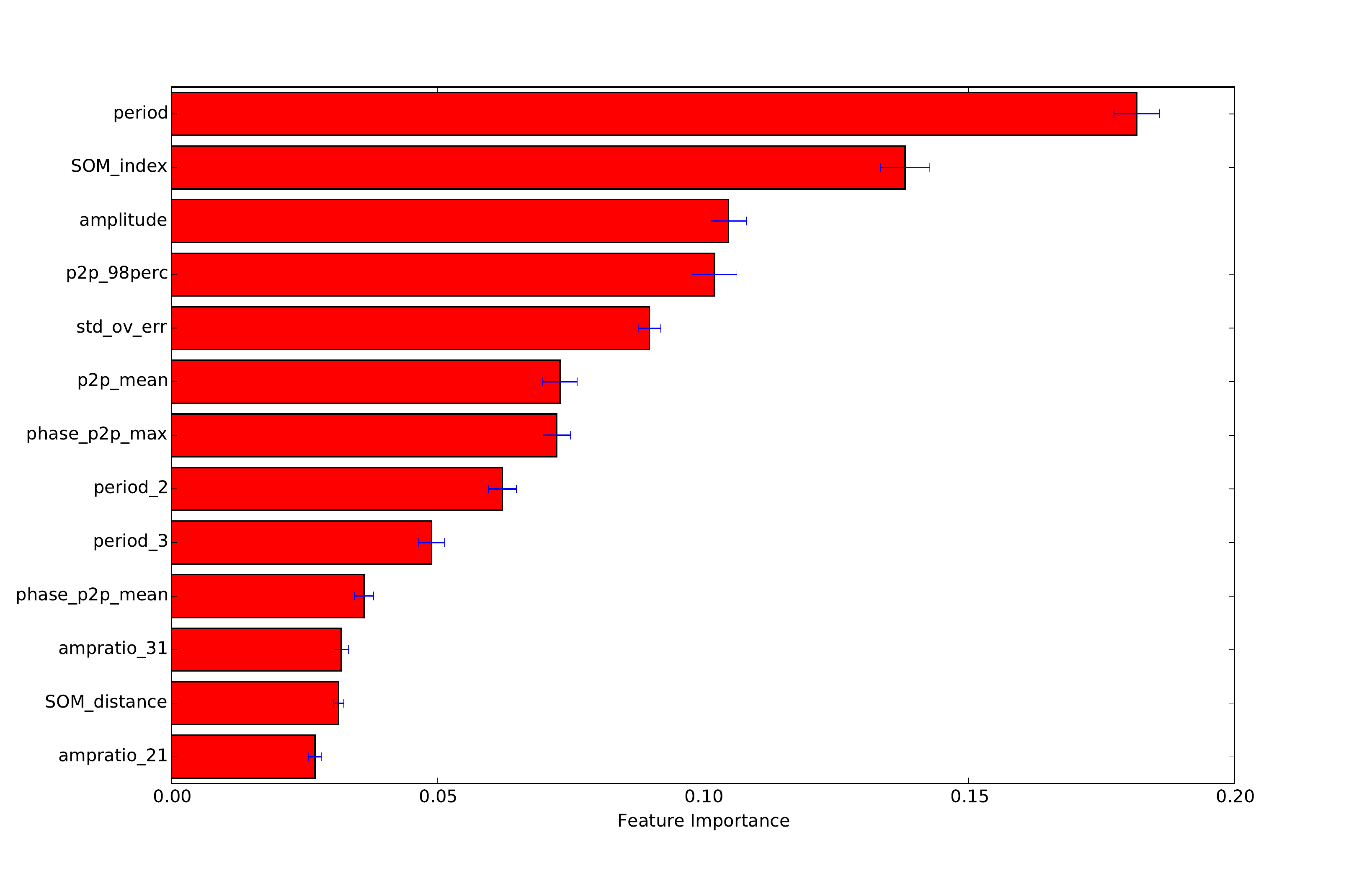}}
\caption{Relative importance of features to the RF. Values and errors arise from the mean and standard deviation of the feature importances extracted from 100 trained classifiers.}
\label{figfeatimportance}
\end{figure}

One advantage of RF classifiers is the ability to estimate feature importance. The classifier naturally measures which features have more descriptive power, through for example how often those features are used in the decision trees, or through the reduction in performance that would be observed is a feature was replaced by a randomly sampled distribution. This allows for model refinement, and is of great use in developing a classifier. We plot the importance of our features in Figure \ref{figfeatimportance}. These are found through training the classifier 100 times, and extracting the mean and standard deviation of the feature importances for each classifier. %The reason for the small changes between classifiers is the random element in each, which arises through selecting subsamples of the training set and features for each tree.

\subsection{Class posterior probability calibration}
\label{sectprobcal}
The RF classifier automatically generates class probabilities (through the proportion of estimators classifying an object into each class). These probabilities are not necessarily accurate. Although it is true that higher class probability means more likelihood of an object being in that class, the probabilities can need calibrating to ensure that they are true posterior probabilities. This is where, if a set of objects have probability p that they are in a certain class, the same proportion of them actually are of that class.

\begin{figure}
\resizebox{\hsize}{!}{\includegraphics{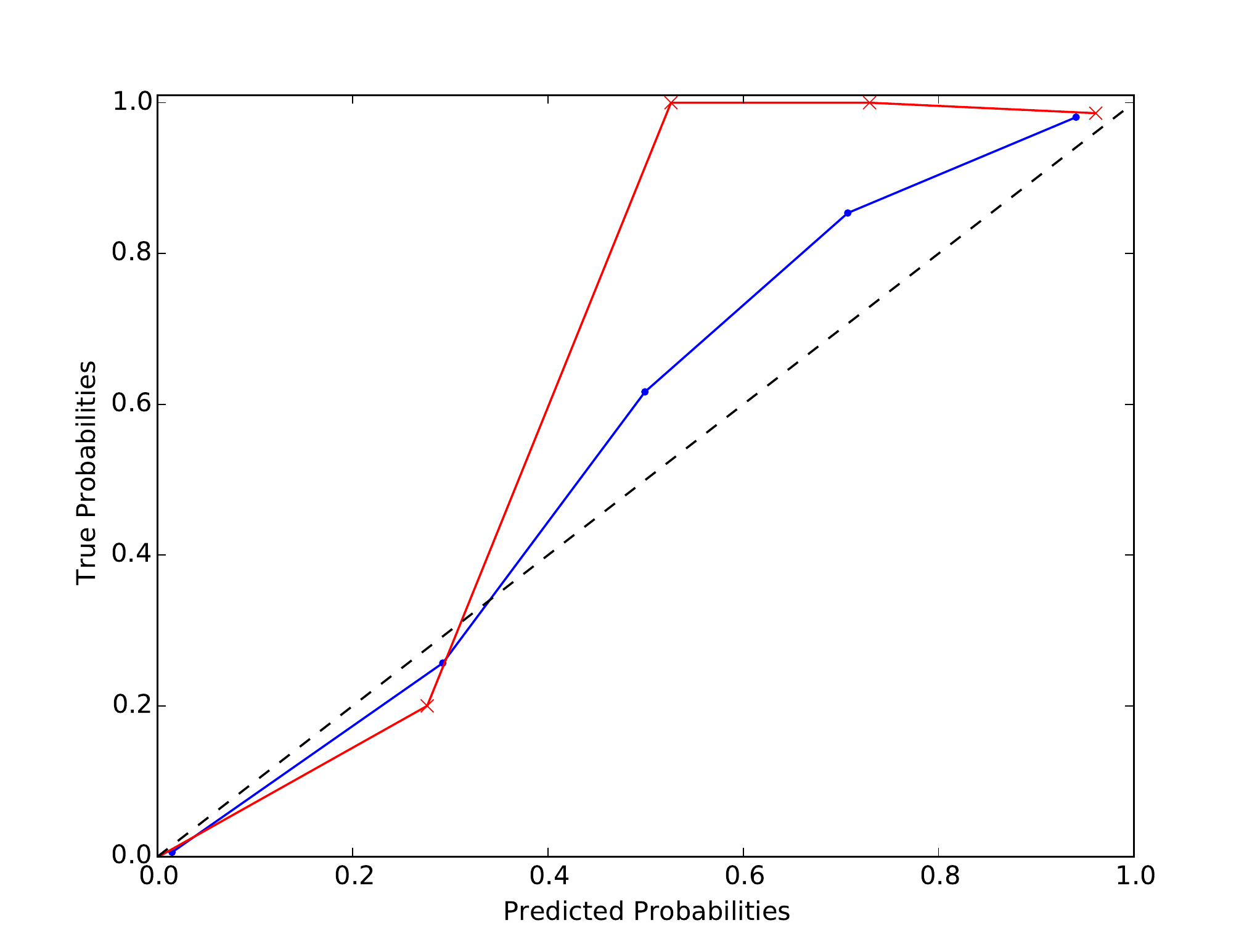}}
\caption{Overall classifier predicted probability against true probability for the RRab class (crosses) and the average of all other classes (dots). The straight black dashed line represents the ideal case.}
\label{figprobcal}
\end{figure}

Initially we test the calibration of our `raw' class probabilities. Figure \ref{figprobcal} shows the class probabilities found from the cross validated training set data created as described in Section \ref{sectRFimplement}. This allows the predicted class probabilities for each training set object to be compared to their known classes. They are clearly not true posterior probabilities, especially for the RRab class, where essentially every object with class probability $>0.5$ is a true class member. For the other classes the given probabilities are closer, but still show some departure from the ideal case.

One common way of testing classifier performance in this way is the Brier score \citep{BRIER:1950hg}. Our raw probabilities have a Brier score of 0.1336. We attempted a number of methods of calibrating them (and so reducing this score). The most usual methods are sigmoid and isotonic regression, which fit certain functions to the calibration curve to transform the probabilities. Similarly to \citet{Richards:2012ea}, we find that these methods are not effective in our case. We attempted the method of \citet{Bostrom:2008dt} to transform the initial class probabilities, but also found the results to be unsatisfactory. Rather than present an incomplete calibration, we give the class probabilities as they are. Users should be aware of this, and avoid interpreting class probabilities as true posterior probabilities.

%. In this scheme, the probabilities undergo the transformation 

%\begin{equation*}
%p_{ic,\textrm{adjusted}} = \begin{cases}
%p_{ic} + r\left(1-p_{ic}\right) & \text{if $p_{ic}$ = max($p_{i1}$,$p_{i2}$,...,$p_{iC}$)}\\
%p_{ic}\left(1-r\right) & \text{otherwise}
%\end{cases}
%\end{equation*}

%\noindent where $p_{ic}$ represents the probability for object $i$ to be in class $c$, C is the number of classes, and r is a scalar between 0 and 1 adjusted to minimise the Brier score. Unlike \citet{Richards:2012ea} we find that this simple transformation is sufficient in our case. We find an optimum value of $r=0.194$, and satisfactory performance even with this simple calibration. The reliability diagram (the proportion of true class objects against their predicted class probabilities, averaged over each of our 6 classes) is shown in Figure \ref{figprobcal}. There are underlying effects not seen in the diagram however. While good overall, the performance on individual classes is more variable. The RRab class is the most extreme example of this, where any object with probability greater than 0.6 tends to be a true RRab. This does not affect classification, only the interpretation of the probabilities.

%relevant brier scores on cv probs: complex: 0.11662 (A=-3,B=1.8), simple: 0.11712, raw: 0.11933

As the training set will not be representative of the true K2 distribution, biases may exist. As the priors are not well known, and the distribution of training sources by no means matches the underlying distribution of variables in K2, true posterior probabilities are impossible to create. Hence the given class probabilities, even if calibrated, would only be posterior probabilities under the assumption that each class has a uniform probability of arising.

\section{Catalogue}

\subsection{Overview}
The full catalogue for K2 fields 0--4 inclusive is given in Table \ref{tabcatalogue}. This Table contains classifications using the Warwick lightcurves, as described in Section \ref{sectExtDet}. The features used to classify these objects are given in Table \ref{tabcatfeatures}.  We also run the classifier on the PDC lightcurves produced by the K2 mission team. These were only available for campaigns 3--4. The resulting classifications are given in Table \ref{tabcatalogueKTeam}, and their associated features in Table \ref{tabcatfeaturesKTeam}.

\begin{landscape}
\begin{table}
\caption{Catalogue table for our Warwick detrended lightcurves. Fields 0--4 are included. Only an extract is shown here for guidance in form. The full table is available online.}
\label{tabcatalogue}
\begin{tabular}{lllllllllll}
\hline
K2 ID & Campaign & Class  & \multicolumn{7}{c}{Class Probabilities} & Anomaly \\
   &        &        &                DSCUT  & EA  & EB  & GDOR & Noise & OTHPER  & RRab &  \\
\hline
202059070 & 0 & Noise & 0.004195 & 0.120507 & 0.016615 & 0.005925 & 0.604636 & 0.246088 & 0.002034 & 0.023891\\
%202059073 & 0 & Noise & 0.000000 & 0.000760 & 0.000037 & 0.000180 & 0.851359 & 0.147664 & 0.000000 & 0.003344\\
% \vdots&\vdots&\vdots&\vdots&\vdots& \vdots &\vdots&\vdots&\vdots&\vdots&\vdots \\
 .&.&.&.&.& . &.&.&.&.&. \\
\hline

\end{tabular}
\end{table}

\begin{table}
\caption{Data features for our Warwick detrended lightcurves. Fields 0--4 are included. Only an extract is shown here for guidance in form. The full table is available online.}
\label{tabcatfeatures}
\begin{tabular}{llllllllllll}
\hline
K2 ID & Campaign & SOM\_index  &  period  & period\_2 & period\_3 & SOM\_distance & phase\_p2p\_mean  &  phase\_p2p\_max  &  amplitude  &  ampratio\_21 & ampratio\_31 \\
  &    &   &  d  & d & d &  & rel. flux & rel. flux & rel. flux & &  \\
\hline
202059070 & 0 & 1544 & 4.764370 & 1.241680 & 0.174448 & 1.180831 & 0.003801 & 0.487419 & 0.042283 & 0.629987 & 0.548721 \\
%202059073 & 0 & 0920 & 8.697272 & 7.072507 & 9.901509 & 0.915877 & 0.004055 & 0.557002 & 0.006845 & 0.937815 & 0.692986 \\
% \vdots&\vdots&\vdots&\vdots&\vdots& \vdots &\vdots&\vdots&\vdots&\vdots&\vdots & \vdots \\
 .&.&.&.&.& . &.&.&.&.&. & . \\
\hline
p2p\_mean  &  p2p\_98perc  &  std\_ov\_err&&&&&&&&& \\
rel. flux & rel. flux &&&&&&&&&& \\
\hline
0.016326 & 0.047548 & 1.310764&&&&&&&&& \\
%0.001782 & 0.005634 & 1.165058&&&&&&&&& \\
 %\vdots&\vdots&\vdots&\vdots&\vdots& \vdots &\vdots&\vdots&\vdots&\vdots&\vdots & \vdots \\
 .&.&.&.&.& . &.&.&.&.&. & . \\
\hline
\end{tabular}
\end{table}

\begin{table}
\caption{Catalogue table for PDC detrended lightcurves. Fields 3--4 only. Only an extract is shown here. The full table is available online.}
\label{tabcatalogueKTeam}
\begin{tabular}{lllllllllll}
\hline
K2 ID & Campaign & Class  & \multicolumn{7}{c}{Class Probabilities} & Anomaly \\
   &        &        &                DSCUT  & EA  & EB  & GDOR & Noise & OTHPER  & RRab &  \\
\hline
205889250 & 3 & Noise & 0.000067 & 0.000000 & 0.000000 & 0.000030 & 0.966544 & 0.033359 & 0.000000 & 0.000000\\
%205890696 & 3 & Noise & 0.000024 & 0.000152 & 0.000000 & 0.000060 & 0.907526 & 0.092238 & 0.000000 & 0.003344\\
 %\vdots&\vdots&\vdots&\vdots&\vdots& \vdots &\vdots&\vdots&\vdots&\vdots\\
  .&.&.&.&.& . &.&.&.&.&. \\
\hline

\end{tabular}
\end{table}

\begin{table}
\caption{Data features for PDC detrended lightcurves. Fields 3--4 only. Only an extract is shown here. The full table is available online.}
\label{tabcatfeaturesKTeam}
\begin{tabular}{llllllllllll}
\hline
K2 ID & Campaign & SOM\_index  &  period  & period\_2 & period\_3 & SOM\_distance & phase\_p2p\_mean  &  phase\_p2p\_max  &  amplitude  &  ampratio\_21 & ampratio\_31 \\
  &    &   &  d  & d & d &  & rel. flux & rel. flux & rel. flux & &  \\
\hline
205889250 & 3 & 0630 & 19.754572 & 12.803889 & 2.281881 & 1.179035 & 0.003795 & 0.421976 & 0.008715 & 0.741302 & 0.592596\\
 %\vdots&\vdots&\vdots&\vdots&\vdots& \vdots &\vdots&\vdots&\vdots&\vdots&\vdots & \vdots \\
 .&.&.&.&.& . &.&.&.&.&. & . \\
\hline
p2p\_mean  &  p2p\_98perc  &  std\_ov\_err&&&&&&&&& \\
rel. flux & rel. flux &&&&&&&&&& \\
\hline
0.005249 & 0.017133 & 1.371857&&&&&&&&& \\
 %\vdots&\vdots&\vdots&\vdots&\vdots& \vdots &\vdots&\vdots&\vdots&\vdots&\vdots & \vdots \\
 .&.&.&.&.& . &.&.&.&.&. & . \\
\hline
\end{tabular}
\end{table}

\end{landscape}

The total number of objects found in each class is given in Table \ref{tabnclass}, at various probability cuts. Note that for RRab class objects in particular, most objects with class probability $>0.5$ are real classifications. In the other cases the probability calibration is better, but these probabilities should still not be interpreted as posterior probabilities.

\begin{table}
\caption{Total objects in each class.}
\label{tabnclass}
\begin{tabular}{lllll}
\hline
Class & Total & Prob $>0.5$ & Prob $> 0.7$ & Prob $> 0.9$ \\
\hline
RRab & 248 & 154 & 72 & 25 \\
DSCUT & 750 & 562 &377 & 166	\\
GDOR & 451 & 264 & 133 & 37	\\
EA & 607 & 308 & 183 & 99 	\\
EB & 463 & 392 & 290 & 186 		\\
OTHPER & 22428 & 18698 & 9399 & 3547 \\
Noise & 43963 & 38609 & 21210 & 6018            \\
\hline
\end{tabular}
\end{table}

We find that the classifier works well on all fields. The RRab class performs well throughout, due to the distinctive shape of their phasecurves. These are well characterised by the SOM. There are however some distinct features unique to fields 3 and 4. The EA class has a tendency to pick up noise dominated lightcurves in these fields, primarily because their point to point scatter is much higher than in fields 0--2. In these cases the class probability, although highest for EA, is still relatively low however. Similarly for DSCUT objects, there are a higher proportion of objects in these fields with many anomalous points, possibly due to flaring or instrumental noise. These points can cause biases in the phase curve, resulting in an artificial sinusoid, which when combined with a short period results in a DSCUT classification. Again these noise objects have a lower probability than real DSCUT lightcurves. One final interesting property is the split between OTHPER and Noise lightcurves. This is good for fields 0--2. In fields 3 and 4, while OTHPER lightcurves are recognised, several Noise lightcurves can be classified as OTHPER. Probability cuts remove the worst of these, but there is no way to distinguish between quasi-periodic instrumental noise and astrophysical variability in this scheme. These issues all lead to the conclusion that the classifier has more trouble with fields 3--4, due to a pattern of increased noise. We expect this issue to improve as K2 detrending methods become more robust.

\subsection{Detrending method comparison}

\begin{table*}
\caption{Total objects in each class in fields 3--4, split by detrending method (W=Warwick, PDC=K2 Team released lightcurves).}
\label{tabdetcomp}
\begin{tabular}{lllllll}
\hline
Class & Total W & Total PDC & Prob $> 0.5$ W & Prob $> 0.5$ PDC &Prob $> 0.7$ W & Prob $> 0.7$ PDC \\
\hline
RRab & 141 & 152 & 95 &115 & 48 & 83  \\
DSCUT & 280 & 266 & 180 & 201 & 116  & 148\\
GDOR & 198 & 382 & 122 & 238 & 61  & 101\\
EA & 255 & 413 & 97 & 223 & 54 & 102 	\\
EB & 168 & 150 & 140 & 131 & 106   & 105	\\
OTHPER & 11402 & 9102 & 8709 & 8034 &  3522  & 4565 \\
Noise & 17143 & 19126 &13012 & 17919 &  3625  & 11566       \\
\hline
\end{tabular}
\end{table*}

Table \ref{tabdetcomp} shows the numbers of variable stars found using each dataset. At first glance the numbers in Table \ref{tabdetcomp} seem to imply significant differences between detrending methods. The discrepancy in RRab numbers is largely a result of differing probability calibration - the same stars are found in both datasets, but those in the Warwick set given lower probabilities (although still higher than all other classes). Other major discrepancies are in the GDOR and EA classes. For GDOR, we find that the PDC set gives better results. Several GDOR lightcurves are misclassified in the Warwick set due to poor detrending masking the true variability. In some cases the PDC GDOR classification is inaccurate, but this is rare for the class probability $>0.7$ objects. For the EA objects, the reverse is true. Several PDC lightcurves are misclassified as EA due to a higher number of lightcurves in the PDC set with very significant remnant outliers. These lead to a high point-to-point scatter, which is interpreted by the classifier as an eclipse. Here the Warwick set is more reliable. The largest absolute difference in the variable classes is in the OTHPER objects, where \mytilde 1000 lightcurves extra pass the high probability cut for the PDC set. This is partly a result of a similar effect as for the RRab objects, where similarly classified objects are given lower probabilities in the Warwick set. However, there are also several objects found in the PDC set which are missed in the Warwick set, due to increased noise levels. The converse is also true, with some lightcurves found in the Warwick set but missed by the PDC. Overall, the two detrending methods perform comparably well, and can be used to reinforce each other when studying variable classes.

\subsection{Anomaly detection}
Due to the limited classification scheme used, it is inevitable that some objects will not fit any of the given classes \citep{Protopapas:2006br}. Due to the inclusion of Noise and OTHPER as classes, this is not a large problem as each class is quite broad. However it is worth noting any particular anomalies. One way of doing this is already intrinsic to the SOM -- the Euclidean distance of a phase curve to its nearest matching pixel template. However this metric only works for periodic sources, and can flag high for noisy sources. We perform a check for anomalies following the method of \citet{Richards:2012ea}. This works by extracting the proximity measure, $\rho_{ij}$ between each tested object $i$ and each object $j$ in the training set. The proximity measure is the proportion of trees in the classifier for which each object ends at the same final classification. It is close to unity for similar objects, and close to zero for dissimilar ones. From the proximity the discrepancy $d$ is calculated, via

\begin{equation}
d_{ij} = \frac{1-\rho_{ij}}{\rho_{ij}}
\end{equation}

The anomaly score is then given by the second smallest discrepancy of an object to the training set. High anomaly scores represent objects which are not well explained by any object in the training set, and are hence outliers.

We find that in this case, the highest few percentiles of anomalous objects are a mixture of noise-dominated lightcurves, unusual eclipsing binaries and variability which does not fit into the used classification scheme. We leave a full analysis of these unusual lightcurves to future work.

\subsection{Eclipsing Binaries}
Encouragingly we identify 139 (96 at class probability $>0.7$) of the 165 EPIC, non-M35 eclipsing binaries identified by \citet{LaCourse:2015jr} in field 0 as either 'EA' or 'EB' type, despite automating the process and not focusing on exclusively eclipsing binaries. The majority of the remainder are identified as 'OTHPER' or 'DSCUT', and are discussed below. We further identify an additional 61 EPIC, non-M35 objects in field 0 as 'EA' or 'EB' at class probability $>0.7$, although as our identification is automated rather than visual some of these may be misidentified by the classifier. Many more eclipsing binaries are found in the other fields.

The previously labelled, but not identified by our classifier, eclipsing binaries fall into three main groups. The first show near-sinusoidal short period lightcurves, and are generally identified as 'DSCUT'. In these cases it is difficult to reliably assign a class with the information available. These objects may be actual $\delta$ Scuti stars, or contact eclipsing binaries. The other and largest group, with 14 members, are identified as 'OTHPER', and show pulsations or spot-modulation in addition to the known eclipses. We note that the classifier will assign a class based largely on the dominant period and phasecurve at this period, hence performs as expected in these cases. Pulsating stars in eclipsing binaries are useful objects, and so while a detailed study of these objects is beyond the scope of this paper we provide a list of such objects in Table \ref{tabqpebs}. These are eclipsing binaries identified by a visual check of the lightcurves performed ourselves (as the \citet{LaCourse:2015jr} catalogue only covered field 0), which are classified as `OTHPER' by our classifier. Some may be blended signals, and hence the pulsator or spot-modulated star may not be a member of the eclipsing binary system.

\begin{table}
\caption{EPIC IDs for 29 visually identified eclipsing binaries classified as `OTHPER' by our classifier, from fields 0--4.}
\label{tabqpebs}
\begin{tabular}{lllll}
\hline
201158453 &201173390  &201569483 & 201584594 \\
  201638314 &202072962 &202137580 & 203371239\\
203476597 &203637922 &204043888 &204193529\\
204328391 &204411840 &205510143 &205919993 \\
 205985357 &205990339  &
206047297 &
206060972 \\
206066862 &
206226010 &
206311743 &
206500801 \\
210350446 &
210501149 &
210766835 &
210945342 \\
211093684 &
211135350 &
  \\
\hline
\end{tabular}
\end{table}

\subsection{$\delta$ Scuti Stars}
\begin{figure}
\resizebox{\hsize}{!}{\includegraphics{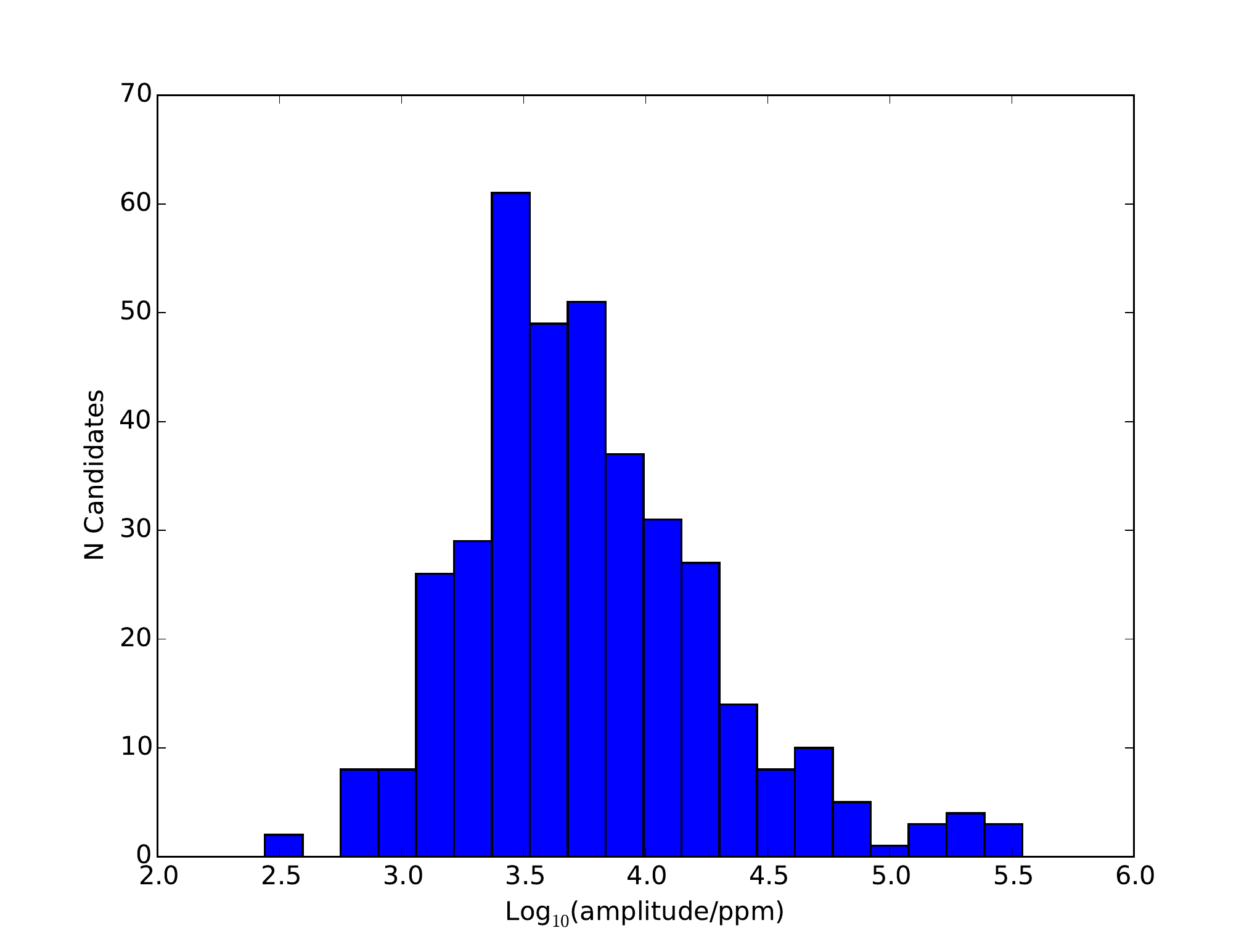}}
\caption{The distribution of phase curve amplitude for DSCUT classified objects. Several high amplitude candidates are visible.}
\label{figdscutampdist}
\end{figure}

\begin{figure}
\resizebox{\hsize}{!}{\includegraphics{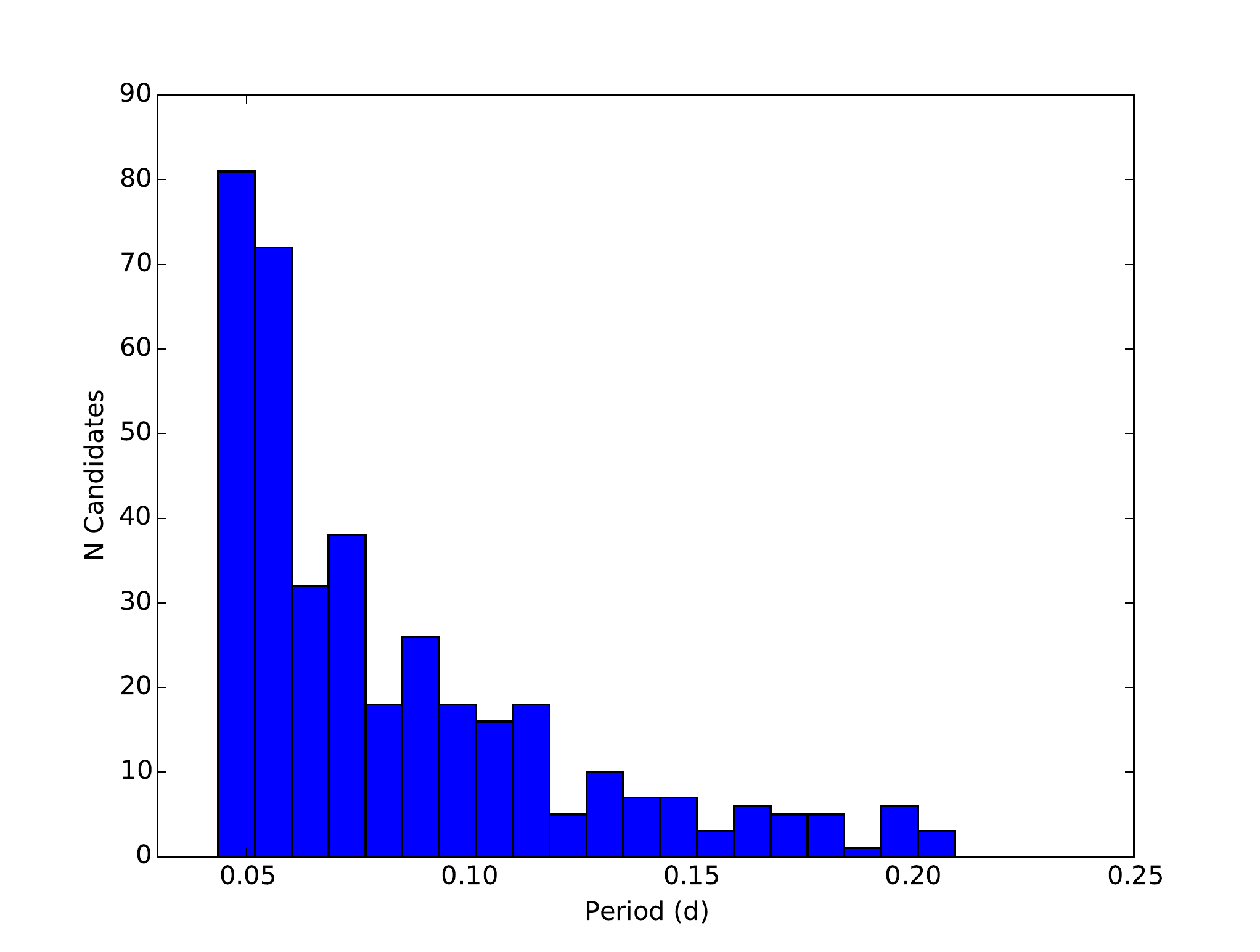}}
\caption{The distribution of pulsation periods for DSCUT classified objects. The cutoff at the low period end is imposed by our Nyquist sampling frequency.}
\label{figdscutperdist}
\end{figure}

We have a sample of 377 $\delta$ Scuti candidates, using a class probability cut of 0.7. The majority of these candidates were previously unknown. It is interesting to study their frequency and amplitude distribution. Note that here we use amplitude defined as in the max-min of the binned phase curve, and semi-amplitude as half this value. The distribution of amplitudes for the 377 $\delta$ Scuti candidates is shown in Figure \ref{figdscutampdist}. We see a number of HADS (high amplitude $\delta$ Scutis). Using an amplitude threshold of $10^4$ ppm as used by \citet{Bradley:2015ep}, 104 of our candidates are HADS. Included in this sample are 11 candidates with an amplitude greater than $10^5$ ppm. The period distribution of the whole sample is shown in Figure \ref{figdscutperdist}, and covers the expected range for $\delta$ Scuti variables, limited by our Nyquist sampling frequency.

As has been mentioned, the DSCUT classified objects are degenerate with $\beta$ Ceph variables due to the lack of colour information available. There is a catalogue of estimated K2 temperatures available for some objects \citep{Stassun:2014wz} which could be used to make probable distinctions if necessary.

\subsection{$\gamma$ Doradus Stars}
\begin{figure}
\resizebox{\hsize}{!}{\includegraphics{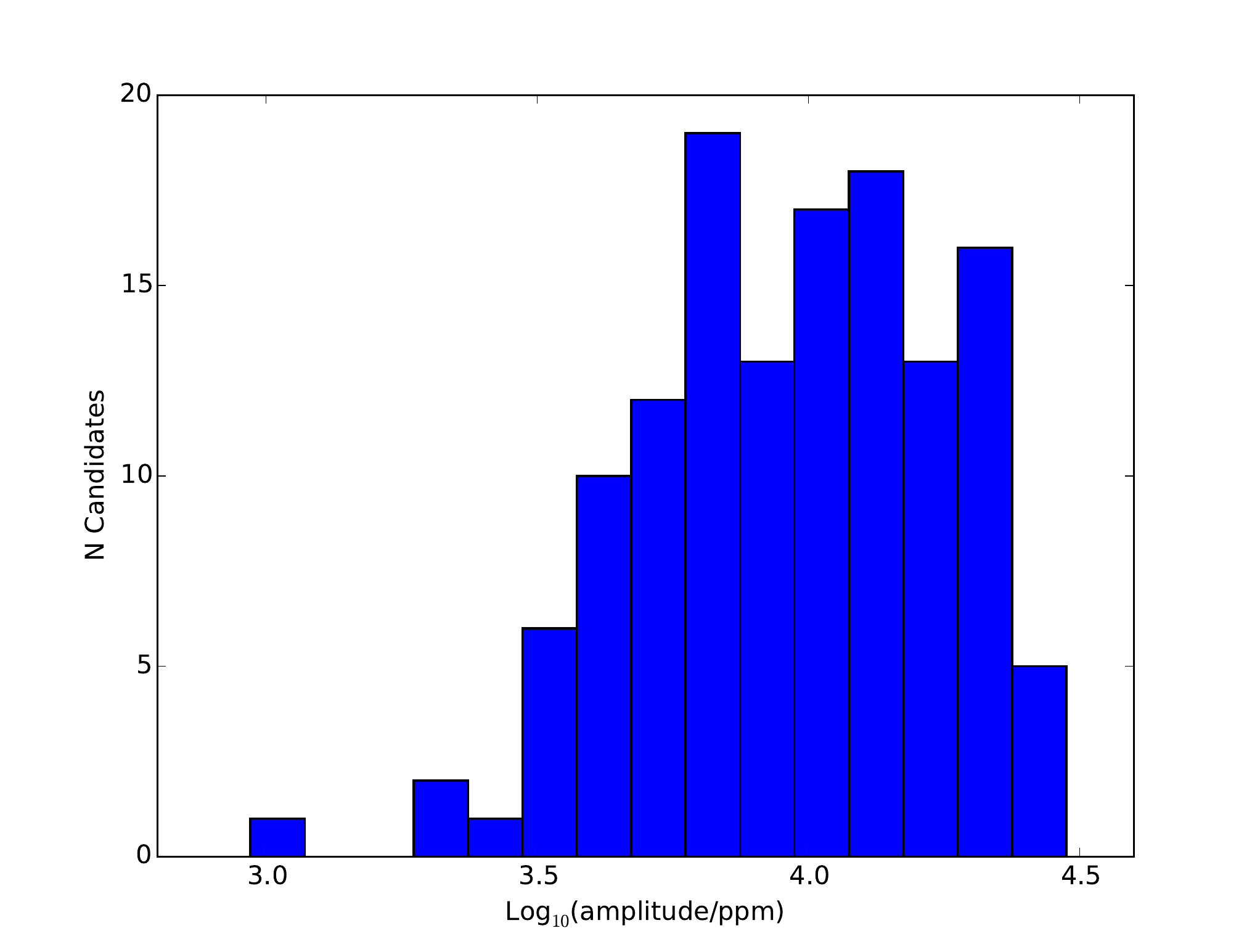}}
\caption{The distribution of phase curve amplitude for GDOR classified objects.}
\label{figgdorampdist}
\end{figure}

\begin{figure}
\resizebox{\hsize}{!}{\includegraphics{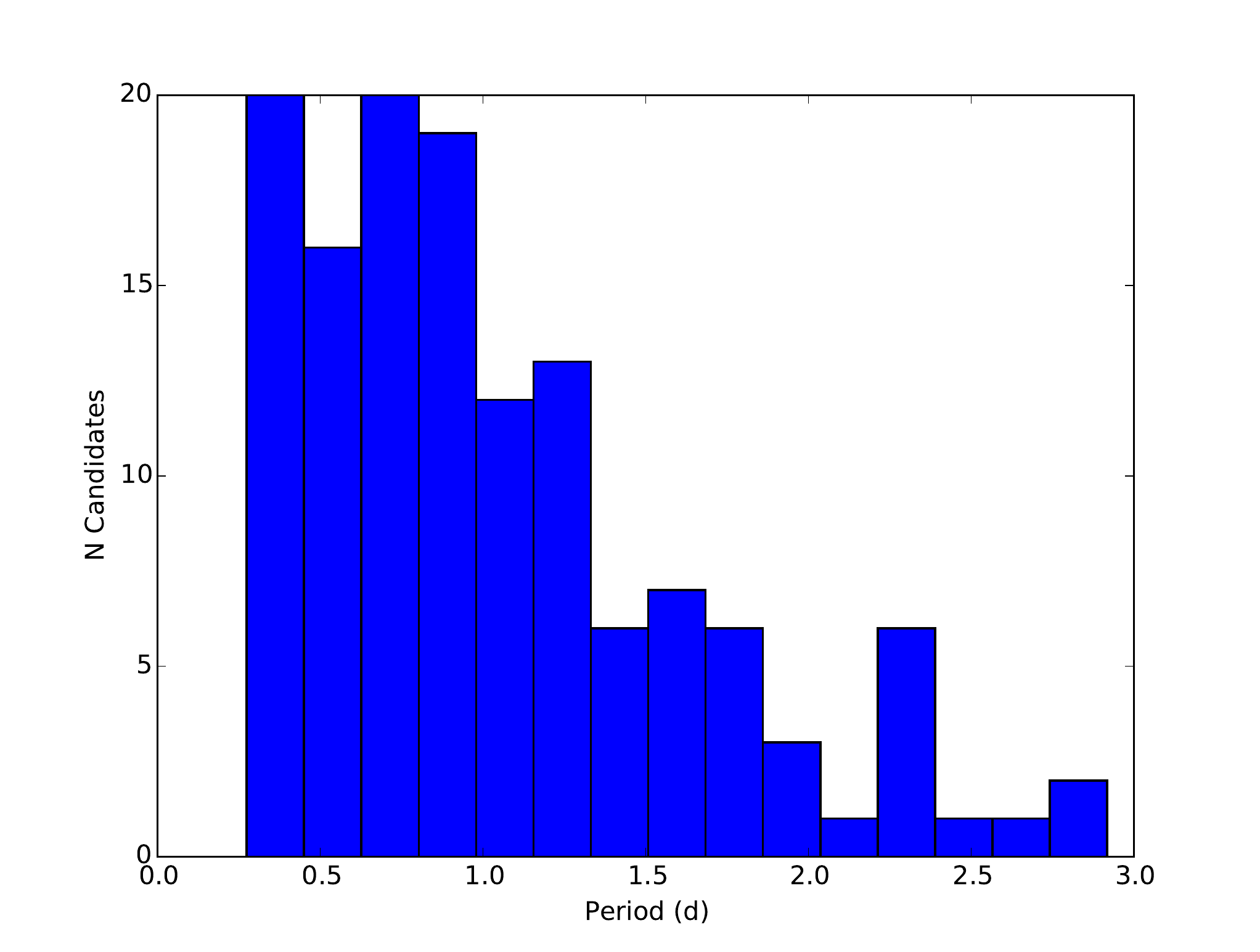}}
\caption{The distribution of pulsation periods for GDOR classified objects.}
\label{figgdorperdist}
\end{figure}

We have a sample of 133 $\gamma$ Doradus candidates, using a class probability cut of 0.7. We plot the amplitude and period distributions in Figures \ref{figgdorampdist} and \ref{figgdorperdist}, following the same definition of amplitude as for the $\delta$ Scuti sample. Note that this amplitude is only for the dominant period phase curve, and so does not include the other significant frequencies often present in $\gamma$ Doradus lightcurves. The period distribution covers the expected range for $\gamma$ Doradus variables. Due to the lack of colour information available, $\gamma$ Doradus objects are degenerate with slowly pulsating B stars.

\subsection{RR Lyrae ab-type Stars}
\begin{figure}
\resizebox{\hsize}{!}{\includegraphics{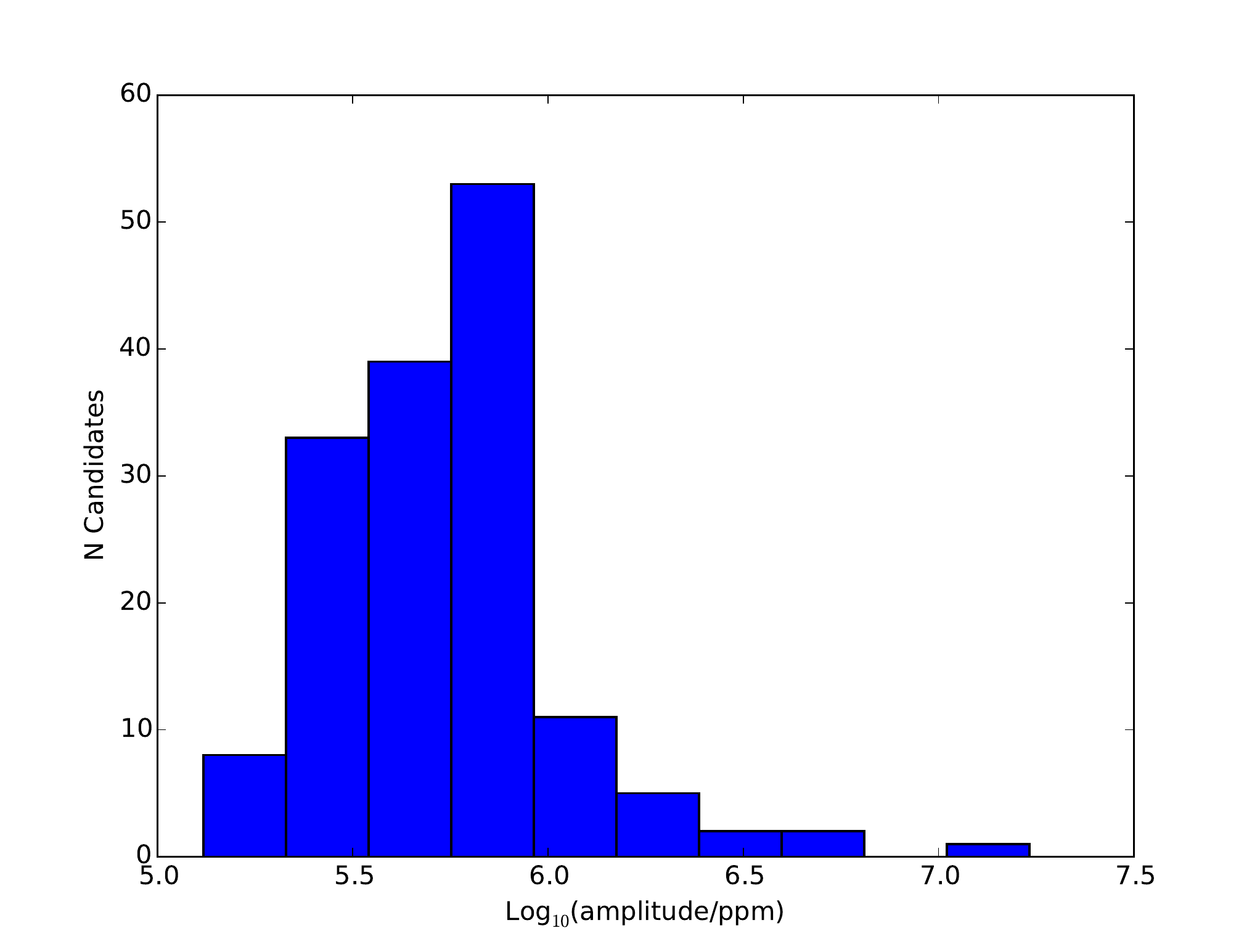}}
\caption{The distribution of phase curve amplitude for RRab classified objects.}
\label{figrrabampdist}
\end{figure}

\begin{figure}
\resizebox{\hsize}{!}{\includegraphics{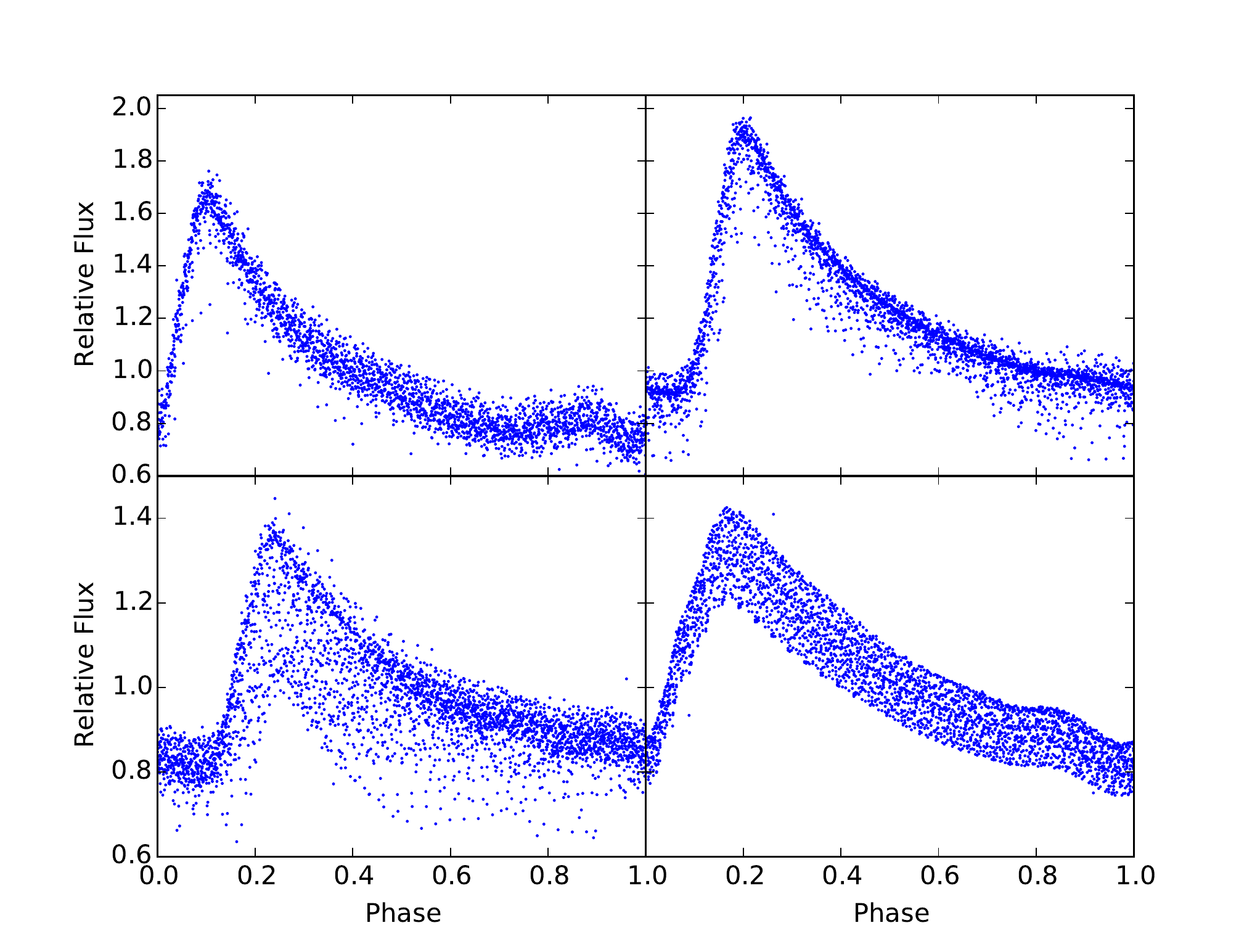}}
\caption{Four phase folded RRAB classified lightcurves. Clockwise from top-left, the EPIC IDs are 210830646, 206409426, 211069540 and 203692906.}
\label{fignewrrab}
\end{figure}

As the RRab class has less well calibrated probability (almost all candidates with Prob(RRab) greater than 0.5 seem to be real) we use an adjusted class probability threshold of 0.5 to study this class. This leaves 154 candidates. Their amplitude distribution is shown in Figure \ref{figrrabampdist}, and peaks at significantly higher amplitude than that of the DSCUT and GDOR candidates as would be expected. Most of these candidates are previously known; we find that 129 of them are in K2 proposals focused on RR Lyrae stars. These proposals contain both known and candidate RR Lyraes; in the candidate cases our classification provides some support for them truly being RR Lyrae variables. Assuming these proposals were comprehensive (reasonable, given the multiple teams involved), this leaves 25 candidates as potential new discoveries by this catalogue. However, as these objects are those not in the proposals, there is a selection effect in favour of misclassified non-RR Lyrae objects. We performed a visual examination of each of these 25 lightcurves, which resulted in 8 of the 25 being confirmed as real RR Lyrae candidates (the others being either misclassified outbursting stars or particularly high amplitude noise). An additional 3 candidates were found by using the PDC lightcurve set and checking objects in both sets with class probability between 0.4 and 0.5, resulting in 10 total new candidates. These objects may still be blends of true RR Lyraes, hence the candidate designation. We plot the phase folded lightcurves for two new discoveries and two known RR Lyrae stars in Figure \ref{fignewrrab}. Some amplitude modulation can be seen, due to some of these targets exhibiting the Blazhko effect \citep{1907AN....175..325B}. RR Lyraes are immensely useful objects, allowing studies of the evolution of stellar populations throughout the Galaxy and in other nearby galaxies. Due to an absolute magnitude-metallicity relation \citep{Sandage:1981ja} it is possible to use them for distance estimation.

\section{Conclusion}
We have implemented a novel combined machine learning algorithm, using both Self Organising Maps and Random Forests to classify variable stars in the K2 data. We consider fields 0--4, and intend to update the catalogue as more fields are released. As more data builds up, it may become possible to implement new variability classes, and study the effect of different detrending methods on the catalogue performance. We obtain a success rate of 92\% using out of bag estimates on the training set.

We train the classifier on a set of Kepler and some K2 data from fields 0--2. As such it is applied completely independently to the majority of the K2 data, and the whole of fields 3--4. That we obtain good results for fields 3--4 bodes well for application of the classifier to future data. 

Algorithms like this will become an increasingly important step in processing the data volumes expected from future astronomical surveys. To maximise scientific return it is critical to select interesting candidates, and do so rapidly and with minimal input. We hope that this method will contribute to the growing body of work attempting to address this issue. 

\section*{Acknowledgements}
The authors thank the anonymous referee for a helpful review of the manuscript. This paper includes data collected by the Kepler mission. Funding for the Kepler mission is provided by the NASA Science Mission directorate. The data presented in this paper were obtained from the Mikulski Archive for Space Telescopes (MAST). STScI is operated by the Association of Universities for Research in Astronomy, Inc., under NASA contract NAS5-26555. Support for MAST for non-HST data is provided by the NASA Office of Space Science via grant NNX13AC07G and by other grants and contracts. We acknowledge with thanks the variable star observations from the AAVSO International Database contributed by observers worldwide and used in this research.

\bibliography{papers190915}
\bibliographystyle{mn2e_fix}

\end{document}